\documentclass[twoside,twocolumn,english,prb,floatfix,showpacs,preprintnumbers,superscriptaddress,amsmath,amssymb]{revtex4}
\usepackage[T1]{fontenc}
\usepackage[latin1]{inputenc}
\usepackage{amsmath}
\usepackage{graphicx}

\makeatletter

\providecommand{\LyX}{L\kern-.1667em\lower.25em\hbox{Y}\kern-.125emX\@}

\usepackage{babel}
\makeatother
\begin{document}

\title{Theoretical analysis of the conductance histograms and structural
properties of \\
Ag, Pt and Ni nanocontacts}

\author{F.~Pauly}

\email{Fabian.Pauly@tfp.uni-karlsruhe.de}

\affiliation{Institut für Theoretische Festk\"{o}rperphysik, Universit\"{a}t
Karlsruhe, D-76128 Karlsruhe, Germany}

\affiliation{Forschungszentrum Karlsruhe, Institut für Nanotechnologie, D-76021
Karlsruhe, Germany}

\author{M.~Dreher}

\affiliation{Fachbereich Physik, Universit\"{a}t Konstanz, D-78457 Konstanz,
Germany}

\author{J.K.~Viljas}

\affiliation{Institut für Theoretische Festk\"{o}rperphysik, Universit\"{a}t
Karlsruhe, D-76128 Karlsruhe, Germany}

\affiliation{Forschungszentrum Karlsruhe, Institut für Nanotechnologie, D-76021
Karlsruhe, Germany}

\author{M.~H\"{a}fner}

\affiliation{Institut für Theoretische Festk\"{o}rperphysik, Universit\"{a}t
Karlsruhe, D-76128 Karlsruhe, Germany}

\affiliation{Forschungszentrum Karlsruhe, Institut für Nanotechnologie, D-76021
Karlsruhe, Germany}

\author{J.C.~Cuevas}

\affiliation{Institut für Theoretische Festk\"{o}rperphysik, Universit\"{a}t
Karlsruhe, D-76128 Karlsruhe, Germany}

\affiliation{Forschungszentrum Karlsruhe, Institut für Nanotechnologie, D-76021
Karlsruhe, Germany}

\affiliation{Departamento de F\'{\i}sica Te\'{o}rica de la Materia Condensada
C-V, Universidad Aut\'{o}noma de Madrid, E-28049 Madrid, Spain}

\author{P.~Nielaba}

\affiliation{Fachbereich Physik, Universit\"{a}t Konstanz, D-78457 Konstanz,
Germany}

\begin{abstract}
Conductance histograms are a valuable tool to study the intrinsic
conduction properties of metallic atomic-sized contacts. These histograms
show a peak structure, which is characteristic of the type of metal
under investigation. Despite the enormous progress in the understanding
of the electronic transport in metallic nanowires, the origin of this
peak structure is still a basic open problem. In the present work
we tackle this issue, extending our theoretical analysis of Au conductance
histograms {[}Dreher \textit{et al.}, PRB \textbf{72}, 075435 (2005){]}
to different types of metals, namely, Ag, Pt and ferromagnetic Ni.
We combine classical molecular dynamics simulations of the breaking
of nanocontacts with conductance calculations based on a tight-binding
model. This combination gives us access to crucial information such
as contact geometries, strain forces, minimum cross-sections, the
conductance, transmissions of the individual conduction channels and,
in the case of Ni, the spin polarization of the current. We shall
also briefly discuss investigations of Al atomic-sized contacts. From
our analysis we conclude that the differences in the histograms of
these metals are due to (i) the very different electronic structures,
which means different atomic orbitals contributing to the transport,
and (ii) the different mechanical properties, which in a case like
Pt lead to the formation of special structures, namely monoatomic
chains. Of particular interest are results for Ni that indicate the
absence of any conductance quantization, and show how the current
polarization evolves (including large fluctuations) from negative
values in thick contacts to even positive values in the tunneling
regime after rupture of the contact. Finally, we also present a detailed
analysis of the breaking forces of these metallic contacts, which
are compared to the forces predicted from bulk considerations.
\end{abstract}


\maketitle

\section{Introduction}

The transport properties and mechanical characteristics of metallic
atomic-scale wires have been the subject of numerous studies over
the past years.\cite{Agrait2003} The analysis of these nanocontacts
is nowadays possible due to experimental techniques like the scanning
tunneling microscope\cite{Agrait1993,Pascual1993} and mechanically
controlled break junctions.\cite{Muller1992} In both cases a metallic
contact is stretched with a precision of a few picometers by the use
of piezoelectric elements, providing very detailed information about
the formation and breaking of metallic systems at the nanoscale.

The relative simplicity of these nanowires makes them ideal systems
to perform extensive comparisons with microscopic theories. Such comparisons
have allowed, in particular, to elucidate the nature of electrical
conduction. The conduction in such systems is usually described in
terms of the Landauer formula, according to which the low-temperature
linear conductance of nonmagnetic contacts can be written as $G=G_{0}\sum _{n}T_{n}$,
where the sum runs over all the available conduction channels, $T_{n}$
is the transmission for the $n$th channel and $G_{0}=2e^{2}/h$ is
the quantum of conductance. As was shown in Ref.~\onlinecite{Scheer1997},
the set of transmission coefficients is amenable to measurement in
the case of superconducting materials. Using this possibility it has
been established that the number of channels in a one-atom contact
is determined by the number of valence orbitals of the central atom,
and the transmission of each channel is fixed by the local atomic
environment.\cite{Cuevas1998a,Scheer1998,Cuevas1998b}

The experiments show that in the stretching processes in which these
metallic wires are formed, the conductance evolves in a step-like
manner which changes from realization to realization. In order to
investigate the typical values of the conductance, different authors
introduced conductance histograms, constructed from a large number
of individual conductance curves.\cite{Olesen1995,Krans1995,Gai1996}
These histograms often show a peak structure, which is specific to
the corresponding metal. Thus for instance, for noble metals like
Au and Ag and for alkali metals like sodium, the conductance has a
certain preference to adopt multiples of $G_{0}$. However, for a
large variety of metals, the peaks do not appear at multiples of $G_{0}$
(for a detailed discussion of the conductance histograms, see section
V.D in Ref.~\onlinecite{Agrait2003}). It has become clear that the
peak structure in the conductance histograms must be related to the
interplay between electronic and mechanical properties. This interplay
was nicely illustrated in the first simultaneous measurement of the
conductance and breaking force,\cite{Agrait1995} but the precise
origin of the differences between the various classes of metals remains
to be understood. The solution of this basic open problem is precisely
the central goal of the present work.

The analysis of the characteristic peaks of the conductance histograms
of alkali and noble metals at relatively high temperatures has revealed
the existence of exceptionally stable radii arising from electronic
shell effects for thin wires and atomic shell effects for thicker
wires.\cite{Yanson1999,Yanson2001,Medina2003,Mares2004,Mares2005}
Stable nanowires with thicknesses of several atoms could also be observed
in transmission electron images.\cite{Rodrigues2000,Kondo2000,Oshima2002}
Commonly, the connection between the peaks in the conductance histograms
and the radius of the contacts is established using semiclassical
arguments based on the Sharvin formula or slight variations of it\cite{Torres1994}\begin{equation}
G=G_{0}\left[\left(\frac{k_{F}R}{2}\right)^{2}-\frac{k_{F}R}{2}+...\right],\label{eq:G-Sharvin}\end{equation}
 where $k_{F}$ is the Fermi wave vector and $R$ is the radius of
the wire.\cite{remarkhardwall} Using this type of formula, it was
suggested in Ref.~\onlinecite{Hasmy2001} that peaks found in the
histogram of the minimum cross-section (MCS) of Al contacts would
immediately translate into peaks in the conductance histograms. In
other words, it was suggested that the conductance peaks would just
be a manifestation of the existence of certain particularly stable
contacts.

From the theory side, the analyses of the conductance histograms are
scarce in the literature. Mostly single stretching events have been
investigated at various levels of sophistication.\cite{Landman1990,Sutton1990,Todorov1993,Bratkovsky1995,Todorov1996,Mehrez1997,Brandbyge1997,Buldum1998,Soerensen1998,Nakamura1999,Jelinek2003,daSilva2004}
The analysis of conductance histograms, however, involves the statistical
exploration of many different stretching events. Most such research
is based on free-electron models, where particular nanowire dynamics
are chosen,\cite{Torres1996} but there are practically no fully atomistic
investigations of the conductance histograms. Two such studies have
just recently appeared, where Dreher \textit{et al.}\cite{Dreher2005}
investigated atomic Au contacts and Hasmy \textit{et al.}\cite{Hasmy2005}
studied Al contacts. In particular, in our work (Ref.~\onlinecite{Dreher2005})
we showed that, at least at low temperatures ($4.2$ K), there is
no simple correspondence between the first peaks in the MCS and the
conductance histograms.

In order to elucidate the origin of the peak structure in the conductance
histograms of metallic atomic-sized contacts, we have extended our
theoretical analysis of the Au conductance histogram\cite{Dreher2005}
to several new metals with varying electronic structures in the present
work. In particular, we have studied the cases of Ag, a noble metal,
Pt, a transition metal, and Ni, a ferromagnetic metal. We shall also
briefly comment on our study of Al (an $sp$-like metal). Our theoretical
approach is based on a combination of classical molecular dynamics
(MD) simulations to describe the contact formation and a tight-binding
(TB) model supplemented with a local charge neutrality condition for
the atomistic computation of the conductance. This combination allows
us to obtain detailed information on the mechanical and transport
properties such as contact geometries, strain forces, the MCS, the
conductance, the number and evolution of individual conductance channels
and, in the case of ferromagnetic contacts, the spin polarization
of the current.

Concerning Ag, we find a sharp peak in the conductance histogram at
$1\, G_{0}$. This peak is due to the formation of single-atom contacts
and dimers in the last stages of the breaking of the wires in combination
with the fact that the transport in the noble metal is dominated by
the $s$ orbitals around the Fermi energy. With \textit{single-atom
contacts} we will refer throughout this article to junctions with
a single atom in the narrowest constriction, in short a one-atom chain,
while \textit{dimer} means an atomic chain consisting of two atoms.
In the case of Pt, the first peak is broadened and shifted to a higher
conductance value (above $1\, G_{0}$). This is due to the fact that
in this transition metal the $d$ orbitals play a fundamental role
in the transport, providing extra conduction channels, as compared
to Ag. For Ni wires, we see that the $d$ orbitals contribute decisively
to the electrical conduction for the minority-spin component, providing
several partially open channels even in the last stages of the stretching
process. As a consequence, we do not observe any type of conductance
quantization. With respect to the polarization of the current, we
see that there is a crossover from large negative values for thick
contacts contacts to positive values in the tunneling regime, right
after the rupture of the contact.

From a more general point of view, the ensemble of our results allows
us to conclude that the differences in the peak structure of the conductance
histograms of metallic nanocontacts can be traced back to the following
two ingredients. First, due to the different electronic structure
of the various classes of metals different atomic orbitals contribute
to the transport. These orbitals determine in turn the number of conducting
channels and therefore the conductance values. Thus, for similar structures
a contact of a multivalent metal will have in general a higher conductance
than one of a noble metal. Second, the different mechanical properties
give rise to the formation of certain characteristic structures, which
are finally reflected in the histograms. For instance, the formation
of monoatomic chains in Au or Pt is responsible for the pronounced
last conductance peak.

The rest of this paper is organized as follows. In Sec.~\ref{sec:Theoretical-approach}
we present the details of our method for simulating the stretching
of atomic wires and show how the conductance is subsequently computed.
Studies of Ag, Pt, Al and Ni contacts follow in Secs.~\ref{sec:Ag},
\ref{sec:Pt}, \ref{sec:Al} and \ref{sec:Ni}, respectively. In each
of these sections we first discuss representative examples of the
stretching processes of the nanocontacts. We then turn to the statistical
analysis of the whole set of simulations for the different metals.
This includes a discussion of the histograms of both the MCS and the
conductance as well as an analysis of the mean channel transmission.
Section \ref{sec:Discussion} is devoted to the discussion of the
mechanical properties of the different metals. Finally, we summarize
the main conclusion of this work in Sec.~\ref{sec:Conclusions}.

\section{The theoretical approach\label{sec:Theoretical-approach}}

The goal of this study is the theoretical description of the mechanical
and electrical properties of metallic nanojunctions. For this purpose,
we make use of the approach introduced in our previous work on the
conductance histogram of Au atomic contacts.\cite{Dreher2005} In
order to analyze ferromagnetic Ni contacts, we need to extend our
method to study also spin-dependent metals. Such an extension is presented
below, but we refer the reader also to Ref.~\onlinecite{Dreher2005}
for supplementary information.

Our theoretical method is based on a combination of classical MD simulations
for the determination of the structure and mechanical properties of
the nanowires and conductance calculations based on a TB model. We
proceed to explain these two types of calculations in the next subsections.

\subsection{Structure calculations}

The breaking of metallic nanocontacts is simulated by means of classical
MD simulations. In all our calculations we assume an average temperature
of $4.2$ K, which is maintained in the simulations by means of a
Nos\'{e}-Hoover thermostat. The forces are calculated using semiempirical
potentials derived from effective-medium theory (EMT).\cite{Jacobsen1996,Stoltzebook}
This theory has already been successfully used for simulating nanowires.\cite{Soerensen1998,Bahn2001,RubioB2001}
For the starting configuration of the contacts we choose a perfect
fcc-lattice of $112$ atoms of length $2.65$ nm (Ag), $2.55$ nm
(Pt), $2.64$ nm (Al) and $2.29$ nm (Ni) oriented along the {[}$001${]}
direction ($z$ direction) with a cross-section of $8$ atoms. This
wire is attached at both ends to two slabs that are kept fixed, each
consisting of $288$ atoms. After equilibration, the stretching process
is simulated by separating both slabs symmetrically by a fixed distance
in every time step ($1.4$ fs). Different time evolutions of the nanocontacts
are obtained by providing the $112$ wire atoms with random starting
velocities. The stretching velocity of $2$ m/s is much bigger than
in the experiment, but it is small compared with the speed of sound
in the investigated materials (of more than $2790$ m/s). Thus the
wire can re-equilibrate between successive instabilities, while collective
relaxation processes may be suppressed.\cite{Todorov1993,Todorov1996}

In order to test whether the conductance changes are correlated with
atomic rearrangements in the nanocontact, we calculate the radius
of the MCS perpendicular to the stretching direction as defined by
Bratkovsky \textit{et al}.\cite{Bratkovsky1995}

Finally, during the stretching process, every $1.4$ ps a configuration
is recorded and the strain force of the nanocontact is computed following
Finbow \textit{et al.}\cite{Finbow1997} Every $5.6$ ps the corresponding
conductance is calculated using the method described below.

\subsection{Conductance calculations\label{sub:Conductance-calculations}}

We compute the conductance within the Landauer approach. To calculate
the electronic structure of our atomic contacts a TB model is employed,
which has been successful in describing the important qualitative
features in the transport through metallic nanojunctions.\cite{Cuevas1998a,Cuevas1998b,Dreher2005}
This model is based on the following Hamiltonian written in a nonorthogonal
local basis \begin{equation}
\hat{H}=\sum _{i\alpha ,j\beta ,\sigma }H_{i\alpha ,j\beta ,\sigma }\hat{c}_{i\alpha ,\sigma }^{\dagger }\hat{c}_{j\beta ,\sigma },\label{eq:nonorthogonalH}\end{equation}
 where $i$ and $j$ run over the atomic sites, $\alpha $ and $\beta $
denote different atomic orbitals and $H_{i\alpha ,j\beta ,\sigma }$
are the on-site ($i=j$) or hopping ($i\neq j$) elements, which are
spin-dependent ($\sigma =\uparrow ,\downarrow $) in the case of ferromagnetic
metals like Ni. Additionally, we need the overlap integrals $S_{i\alpha ,j\beta }$
of orbitals at different atomic positions.\cite{remarkSnospin} We
obtain the quantities $H_{i\alpha ,j\beta ,\sigma }$ and $S_{i\alpha ,j\beta }$
from a parameterization that is designed to accurately reproduce the
band structure of bulk materials.\cite{Papaconstantopoulos,Haftel2004}
The atomic basis is formed by $9$ valence orbitals, namely the $s$,
$p$ and $d$ orbitals which give rise to the main bands around the
Fermi energy. In this parameterization both the hoppings and the overlaps
to a neighboring atom depend on the interatomic position, which allows
us to apply this parameterization in combination with the MD simulations.
The overlap and hopping elements have a cutoff radius that encloses
up to $9$ (Ag, Pt and Al) or $12$ (Ni) nearest-neighbor shells.
The left ($L$) and right ($R$) electrodes are constructed such that
all the hopping elements from the $112$ wire atoms, which we will
call the central part or center of our contact ($C$), to the electrodes
are taken into account. This means that the electrodes in the conductance
calculation are constituted of {[}$001${]} layers containing even
more than the $288$ slab atoms used in the structure calculations.
Note that with the word electrode we will refer, throughout this article,
to the fixed slab atoms (or the extended {[}$001${]} layers used
in the conductance calculations). 

The local environment in the neck region is very different from that
in the bulk material for which the TB parameters have been developed.
This can cause large deviations from the approximate local charge
neutrality that typical metallic elements must exhibit. Within the
TB approximation we correct this effect by imposing a local charge
neutrality condition on the atoms in the central part of the nanowire
through a self-consistent variation of the Hamiltonian. This self-consistent
procedure requires the computation of the electronic density matrix
$P_{i\alpha ,j\beta }$, which is obtained by integrating the Green
function of the center up to the Fermi energy\cite{remarkPintegration}\begin{equation}
\hat{P}_{i\alpha ,j\beta }=-\frac{1}{\pi }\int _{-\infty }^{E_{F}}\textrm{Im}\left[\sum _{\sigma }\hat{G}_{CC,\sigma }^{r}\left(E\right)\right]dE.\label{eq:Pmat}\end{equation}
In this expression $\hat{G}_{CC,\sigma }^{r}$ is the retarded Green's
function of the central part of the contact\begin{equation}
\hat{G}_{CC,\sigma }^{r}\left(E\right)=\left[E\hat{S}_{CC}-\hat{H}_{CC,\sigma }-\hat{\Sigma }_{L,\sigma }^{r}-\hat{\Sigma }_{R,\sigma }^{r}\right]^{-1},\label{eq:DefGCCsigma}\end{equation}
 where $\sigma $ stands for the spin component, $\hat{S}_{CC}$ is
the overlap matrix of the center, $\hat{H}_{CC,\sigma }$ is the Hamiltonian
and $\hat{\Sigma }_{X,\sigma }$ (with $X=L$ or $R$) are the self-energies
that describe the coupling of the center to the electrodes. They are
given by \begin{equation}
\hat{\Sigma }_{X,\sigma }^{r}\left(E\right)=\left(\hat{H}_{CX,\sigma }-E\hat{S}_{CX}\right)\hat{g}_{XX,\sigma }^{r}\left(\hat{H}_{XC,\sigma }-E\hat{S}_{XC}\right),\label{eq:SigmaXsigma}\end{equation}
 with the unperturbed retarded electrode Green's function $\hat{g}_{XX,\sigma }^{r}$
and the overlap (hopping) matrices from the center to the electrodes
$\hat{S}_{CX}$ ($\hat{H}_{CX,\sigma }$). The unperturbed electrode
Green's functions are assumed to be bulk Green's functions in all
our calculations. The charge on the atom $i$ is then determined using
a Mulliken population analysis \begin{equation}
N_{i}=\sum _{\alpha }\left(\hat{P}_{CC}\hat{S}_{CC}\right)_{i\alpha ,i\alpha },\label{eq:charge_Mulliken}\end{equation}
where only the contributions of the central part to the atomic charge
are considered.\cite{remarkchargeCC,remarkMulliken} The new Hamiltonian
matrix elements $H_{i\alpha ,j\beta ,\sigma }$ are obtained from
the original ones $H_{i\alpha ,j\beta ,\sigma }^{\left(0\right)}$
as\cite{Brandbyge1999}\begin{equation}
H_{i\alpha ,j\beta ,\sigma }=H_{i\alpha ,j\beta ,\sigma }^{\left(0\right)}+S_{i\alpha ,j\beta }\frac{\phi _{i}+\phi _{j}}{2}.\label{eq:Hn_shift}\end{equation}
 The shifts $\phi _{i}$ are determined such that no atom deviates
from the charge neutrality by more than $0.02$ electron charges ($\left|N_{i}-N_{\textrm{atom}}\right|<0.02$,
and $N_{\textrm{atom}}$ stands for the electronic charge of the respective
charge-neutral metal atom). Note that there is one shift parameter
per central atom, also in the case of ferromagnetic metals.

The low-temperature linear conductance is then computed using a Green's
function formalism (see Ref.~\onlinecite{Dreher2005} for details),
finally resulting in the Landauer formula 

\begin{equation}
G^{\sigma }\left(E_{F}\right)=\frac{e^{2}}{h}\sum _{n}T_{n}^{\sigma }\left(E_{F}\right)\label{eq:conductanceG_spin}\end{equation}
 with the Fermi energy $E_{F}$ and the transmission $T_{n}^{\sigma }$
of the $n$th transmission eigenchannel. The conductance is then given
as the sum over the different spin contributions \begin{equation}
G=\sum _{\sigma }G^{\sigma },\label{eq:conductanceGsumspin}\end{equation}
which has the form \begin{equation}
G=G_{0}T\left(E_{F}\right)=G_{0}\sum _{n}T_{n}\label{eq:conductanceG}\end{equation}
 for the spin independent case. As explained in the introduction,
$G_{0}=2e^{2}/h$ is the quantum of conductance, and $T$ is the total
transmission.

To investigate the influence of a small bias voltage, we have computed
for Ag and Pt the transmission $T\left(E\right)$ in an energy interval
of width $2\Delta =100$ meV around the Fermi energy.\cite{remarkTaverage}
The averaged conductance\begin{equation}
\left\langle G\right\rangle =G_{0}\left\langle T\right\rangle =G_{0}\frac{1}{2\Delta }\int _{E_{F}-\Delta }^{E_{F}+\Delta }T\left(E\right)dE,\label{eq:conductanceGav}\end{equation}
 can then be compared to the conductance $G=G_{0}T\left(E_{F}\right)$
at the Fermi energy (cf.~Eq.~(\ref{eq:conductanceG})). This provides
information on the nonlinearity of current-voltage characteristics,
although the formulas we use are, strictly speaking, only valid for
the zero-bias situation.

\subsection{Local density of states calculations}

To gain some insight into the electronic states relevant for the transport
through our nanowires, we shall also compute the local density of
states (LDOS) projected onto particular atoms. The computation of
the LDOS requires the evaluation of the Green function of the central
part of the nanowire $\hat{G}_{CC,\sigma }$ (cf.~Eq.~(\ref{eq:DefGCCsigma})).
From $\hat{G}_{CC,\sigma }$ we construct the LDOS via a L\"{o}wdin
transformation.\cite{Szabo1996} The LDOS for a particular orbital
$\alpha $ of atom $i$ is then given by\cite{remarketa}\begin{equation}
\textrm{LDOS}_{i\alpha ,\sigma }\left(E\right)=-\frac{1}{\pi }\textrm{Im}\left[\hat{S}_{CC}^{1/2}\hat{G}_{CC,\sigma }^{r}\left(E\right)\hat{S}_{CC}^{1/2}\right]_{i\alpha ,i\alpha }.\label{eq:LDOS_i_alpha_sigma}\end{equation}
In the case of the nonferromagnetic metals (Ag and Pt) the LDOS will
in the following be given only for one spin component, because of
the spin degeneracy.

\section{Silver atomic contacts\label{sec:Ag}}

We start the analysis of our results with the discussion of the conductance
of Ag nanowires. Ag is, like Au, a noble metal with a single valence
electron. Different experiments have shown that the conductance of
Ag contacts exhibits a tendency towards quantized values in the last
stages of the wire formation.\cite{Ludoph2000,AIYansonPhD,Rodrigues2002}
In fact, the most dominant feature in the experimental low-temperature
conductance histogram is a pronounced peak at $1\, G_{0}$.\cite{Ludoph2000,AIYansonPhD}

\subsection{Evolution of individual silver contacts\label{sub:Ag_evolution}}

\begin{figure}
\includegraphics[  scale=0.25,
  keepaspectratio]{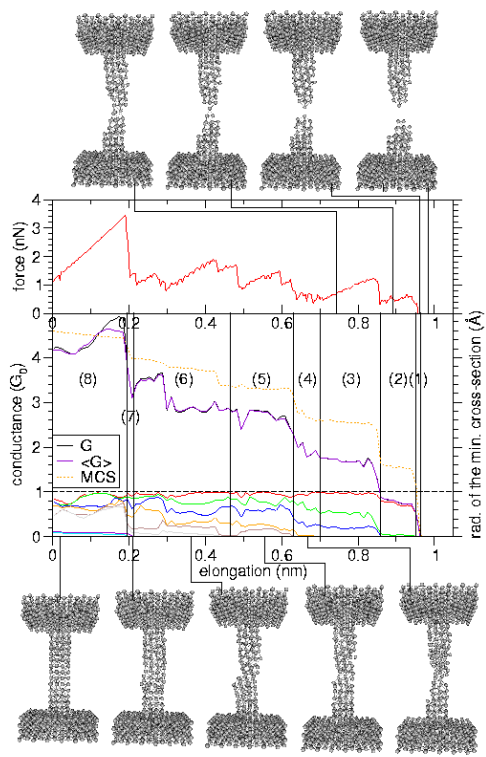}

\caption{\label{cap:Ag-contact1}(Color online) Formation of a single-atom
contact for Ag ($4.2$ K, {[}$001${]} direction). The upper panel
shows the strain force as a function of the elongation of the contact.
In the lower panel the conductance $G$, the averaged conductance
$\left\langle G\right\rangle $, the MCS (minimum cross-section) radius
and the channel transmissions are displayed. Vertical lines separate
regions with different numbers of open channels ranging from $8$
to $1$.\cite{remarkchannelregions} Above and below these graphs
snapshots of the stretching process are shown.}
\end{figure}

Let us first describe some typical examples of the breaking of Ag
nanowires. In Fig.~\ref{cap:Ag-contact1} we show the formation of
a single-atom contact. Beside the strain force we display the conductance
$G$, the averaged conductance $\left\langle G\right\rangle $ (cf.~Eq.~(\ref{eq:conductanceGav})),
the MCS radius and the channel transmissions.%
 As one can see, after an initial evolution up to an elongation of
$0.2$ nm (region with eight conduction channels), which is similar
for all the $50$ Ag contacts studied, the conductance starts decreasing
in a step-like manner which changes from realization to realization.
The jumps in the conductance usually occur at plastic deformations
of the contact, i.e.~when bonds break and sudden atomic rearrangements
take place. Such sudden rearrangements are visible as a break-in of
the strain force. The elastic stages, in which the atomic bonds are
being stretched, are characterized by a linear increase of the strain
force. In this case the conductance exhibits well-defined plateaus
(see for instance the region with three channels, which occurs for
elongations between $0.7$ nm and $0.83$ nm). In the last stages
of the breaking of the contact, displayed in Fig.~\ref{cap:Ag-contact1},
a stable single-atom contact is formed. In this region the conductance
is mainly dominated by a single channel, although a second one is
still visible (see two-channel region or elongations between $0.86$
nm and $0.95$ nm). Subsequently, a dimer structure is formed, which
survives for a short period of time, after which the contact finally
breaks. In this region only a single transmission channel is observed.

It is worth noticing that there is practically no difference between
the conductance $G$ and the averaged conductance $\left\langle G\right\rangle $
(cf.~Eq.~(\ref{eq:conductanceGav})), demonstrating that the transmission
as a function of the energy is rather flat around the Fermi energy
(in the window $-\Delta \leq E-E_{F}\leq \Delta $). This can be seen
explicitly in Fig.~\ref{cap:Ag-T-LDOS-E-contact2}, which we shall
discuss later in more detail. The flat transmission $T(E)$ is expected
for a noble metal like Ag because its density of states around $E_{F}$
is mainly dominated by the contributions of the $s$ and $p$ bands,
which are rather broad and vary slowly with energy.

\begin{figure}
\includegraphics[  scale=0.25,
  keepaspectratio]{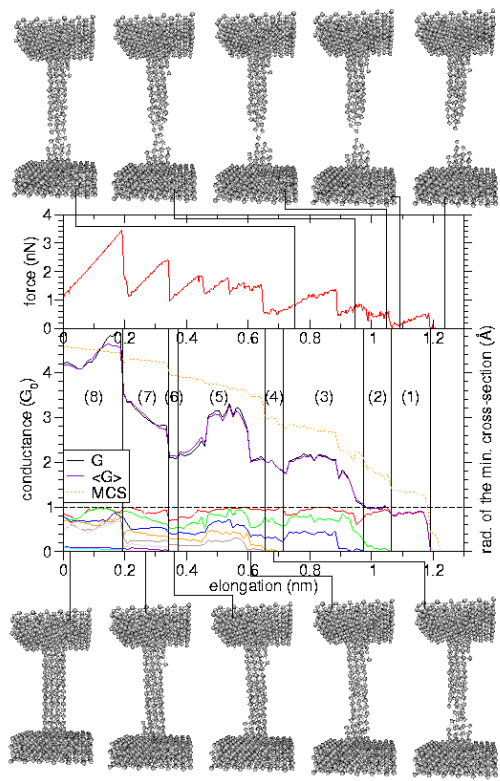}

\caption{\label{cap:Ag-contact2}(Color online) Formation of a dimer configuration
for Ag ($4.2$ K, {[}$001${]} direction). The upper panel shows the
strain force as a function of the elongation of the contact. In the
lower panel the conductance $G$, the averaged conductance $\left\langle G\right\rangle $,
the MCS (minimum cross-section) radius and the channel transmissions
are displayed. Vertical lines separate regions with different numbers
of open channels ranging from $8$ to $1$.\cite{remarkchannelregions}
Above and below these graphs snapshots of the stretching process are
shown.}
\end{figure}

\begin{figure}
\includegraphics[  scale=0.25]{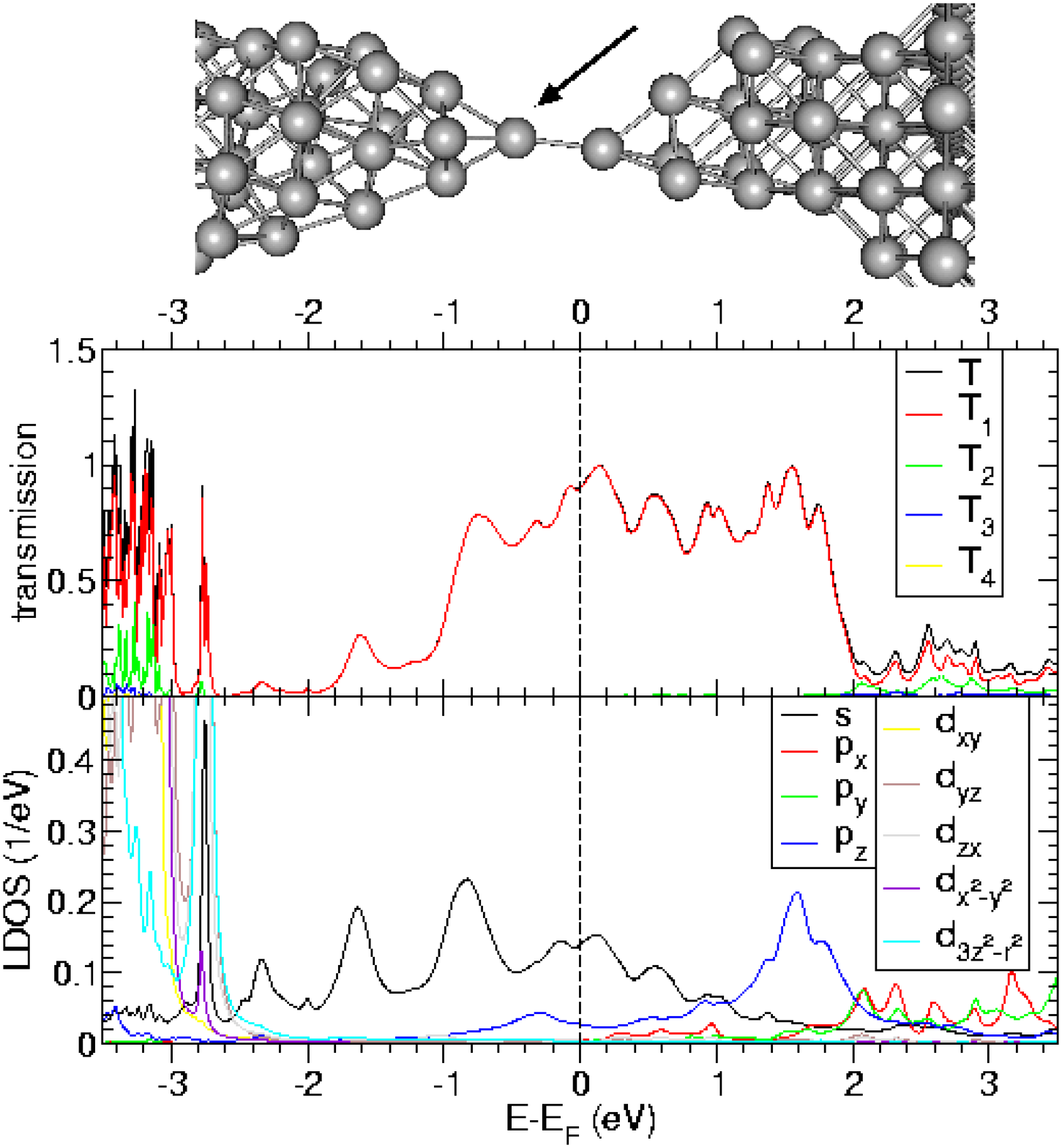}

\caption{\label{cap:Ag-T-LDOS-E-contact2}(Color online) Ag contact of Fig.~\ref{cap:Ag-contact2}
at an elongation of $1.16$ nm. The total transmission $T$ is plotted
as a function of the energy together with the contributions from the
different transmission channels $T_{n}$. Additionally the LDOS (local
density of states) is given for an atom in the narrowest part of the
contact, where the different orbital contributions have been itemized.
Above the figure the narrowest part of the Ag contact is displayed
in a magnified fashion and the atom is indicated, for which the LDOS
is shown.}
\end{figure}

Another example of a breaking curve for Ag is depicted in Fig.~\ref{cap:Ag-contact2}.%
 In the beginning the conductance evolves like for the contact discussed
above (cf.~Fig.~\ref{cap:Ag-contact1}). This time a stable dimer
is finally formed. Prior to the formation of the dimer structure,
which sustains a single channel (see one-channel region or elongations
from $1.06$ nm to $1.19$ nm), there also appears a single-atom contact,
where two channels are still visible (see two-channel region or elongations
from $0.97$ nm to $1.06$ nm), in analogy to what has been found
for Au before.\cite{Dreher2005} We observe for both configurations
a single dominant transmission channel and a conductance of around
$1\, G_{0}$. This result is consistent with first-principles calculations,
where it has been shown for selected ideal contact geometries that
the transmission of Ag chains is around $1\, G_{0}$ and the conductance
is carried by a single transmission channel.\cite{Mozos2002,Lee2004}

Due to the appearance of a stable dimer structure there is now a long
and flat last plateau before rupture in Fig.~\ref{cap:Ag-contact2}.
Our simulations show that this type of dimer is the most common structure
in the last stages of the contact formation.

A certain peculiarity can be observed, if one has a closer look at
the region with six open channels. Here, the conductance first drops
abruptly and then it increases again in the region with five open
channels. Notice that this increase is accompanied by a steady decrease
of the MCS. This type of reentrance of the conductance, which is often
observed experimentally, cannot be explained in terms of semiclassical
arguments, which are based on Eq.~(\ref{eq:G-Sharvin}). According
to this formula the conductance should be a monotonous function of
the MCS, which, however, is not always the case. Such break-ins of
the conductance have already been observed before in simpler TB calculations.\cite{Bratkovsky1995}

In order to explain the existence of a single channel in the final
stages of breaking, we have plotted in Fig.~\ref{cap:Ag-T-LDOS-E-contact2}
the LDOS of an atom in the narrowest part of the junction as a function
of the energy together with the transmission.%
 We have chosen a dimer configuration at an elongation of $1.16$
nm, right before the rupture of the contact displayed in Fig.~\ref{cap:Ag-contact2}.
The transmission around the Fermi energy is made up of a single channel,
exhibiting only a tiny variation in the energy window $-\Delta \leq E-E_{F}\leq \Delta $.
In the LDOS there are two dominant contributions, one coming from
the $s$ orbital, as expected, and the other one from the $p_{z}$
orbital. Therefore, the transmission channel is expected to consist
mainly of these two contributions, the other orbitals being of minor
importance. As found before,\cite{Cuevas1998b,Brandbyge1999} the
$s$ and $p_{z}$ orbitals are then forming a radially isotropic transmission
channel along the transport direction. If we denote by $l_{z}$ the
projection of the angular momentum onto the $z$ axis (transport direction),
this channel has the quantum number $l_{z}=0$.

\subsection{Statistical analysis of silver contacts}

\begin{figure}
\includegraphics[  scale=0.25,
  keepaspectratio]{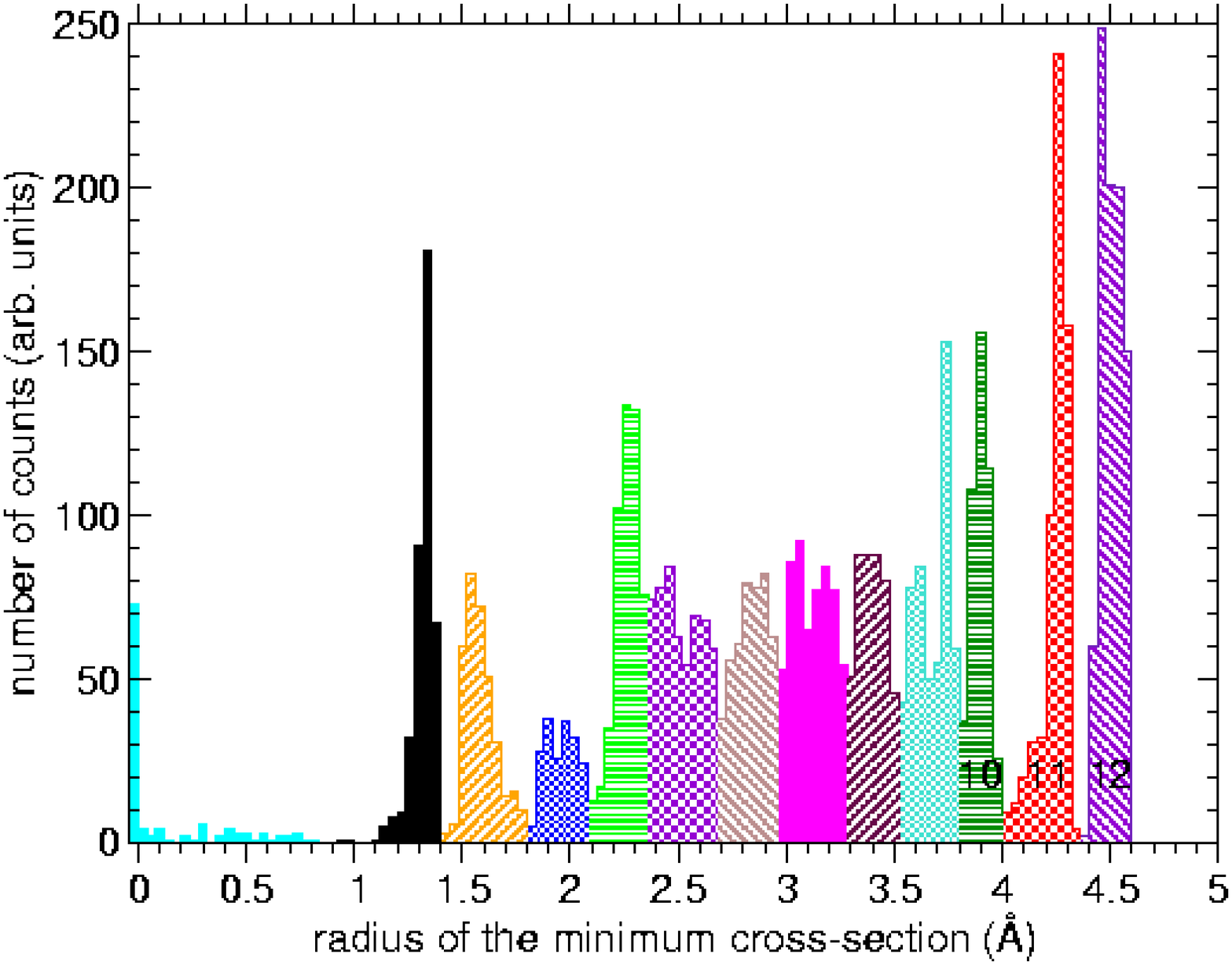}

\vspace{0.1cm}
\includegraphics[  scale=0.25,
  keepaspectratio]{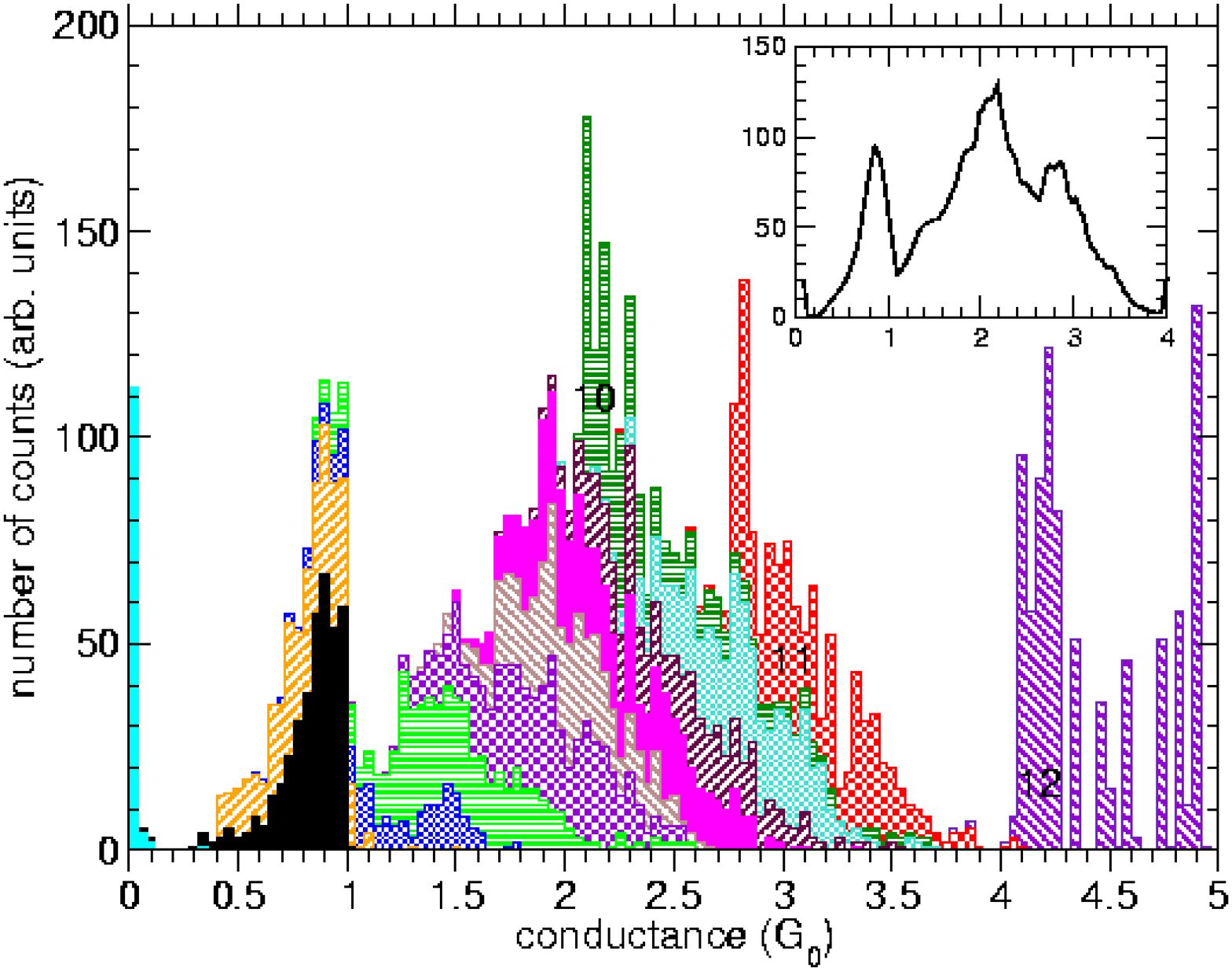}

\caption{\label{cap:Ag-mcs-condhisto}(Color online) MCS (minimum cross-section)
histogram (upper panel) and conductance histogram (lower panel) for
Ag ($4.2$ K, {[}$001${]} direction, $50$ contacts). In the MCS
histogram different regions of frequently occurring radii have been
defined with different pattern styles. The patterns in the conductance
histogram indicate the number of counts for conductances belonging
to the corresponding region of the MCS histogram. For better reference
in the text, some regions in the MCS and conductance histogram have
additionally been labeled with numbers. In the inset of the lower
panel the conductance histogram is displayed in the relevant region
in a smoothed version by averaging over six nearest-neighbor points. }
\end{figure}

In Fig.~\ref{cap:Ag-mcs-condhisto} our computed MCS histogram as
well as the computed conductance histogram are displayed.%
 The histograms are obtained by collecting the results of the stretching
of $50$ Ag contacts oriented along the {[}$001${]} direction at
$4.2$ K, as described in Sec.~\ref{sec:Theoretical-approach}. In
the case of the MCS histogram, the most remarkable feature is the
appearance of very pronounced peaks, which indicate the existence
of particularly stable contact radii. For the purpose of correlating
these peaks with the structure in the conductance histogram, we have
marked the regions around the peaks in the MCS histogram with different
pattern styles. In the conductance histogram we indicate the counts
for conductances belonging to a certain MCS region with the same pattern
style, in order to establish this correlation between the geometric
structure of the contacts and the features in the conductance histogram.

With respect to the conductance histogram, our main result is the
appearance of a pronounced peak at $1\, G_{0}$, in accordance with
the experimental results.\cite{Ludoph2000,AIYansonPhD,Rodrigues2002}
This peak mainly stems from the contributions of contacts with MCS
radii in the first (dimers) and second (single-atom contacts) region
of the MCS histogram. Therefore, we can conclude that the peak at
$1\, G_{0}$ is a consequence or manifestation of the formation of
single-atom contacts and dimers in the last stages of the breaking
of the Ag wires.

It is also important to stress that the contributions to the conductance
histogram coming from different regions of the MCS histogram clearly
overlap. This means in practice that the MCS radius is not the only
ingredient that determines the conductance, as one would conclude
from semiclassical arguments (see Eq.~(\ref{eq:G-Sharvin})). In
other words, the peak structure in the MCS histogram is not simply
translated into a peak structure in the low-temperature conductance
histogram, as suggested in Ref.~\onlinecite{Hasmy2001}.

At this stage, a word of caution is pertinent. In break junction experiments,
contacts are opened and closed repeatedly, and the breaking processes
starts with a conductance as large as $100\, G_{0}$.\cite{Mares2005}
Compared to this value, our simulations start with a very small conductance
of around $4\, G_{0}$. Additionally, all the contacts are oriented
along the {[}$001${]} direction, which can be expected to have an
influence on the structure of the conductance histogram. Even for
rather thick contacts it has been shown experimentally that prefabricated
wires cause a different peak structure in the conductance histograms.\cite{Yanson2005}

The last three peaks of the MCS histogram (labeled $10$, $11$ and
$12$ in Fig.~\ref{cap:Ag-mcs-condhisto}) are mainly dominated by
the (arbitrarily) selected {[}$001${]} starting-configuration. It
is interesting to observe that the MCS region labeled with a $10$
has a large weight at conductances of somewhat above $2\, G_{0}$,
although it should be expected to have contributions for large transmissions
because of its high MCS. The break-in of the conductance in Fig.~\ref{cap:Ag-contact2}
at the transition from the six- to five-channel region is an example
showing the origin of the large weight of this MCS region at $2\, G_{0}$.
This observation illustrates that even conductance regions down to
$2\, G_{0}$ are distorted due to the small size of our contacts.
While we can be sure about the first peak in the conductance histogram
at $1\, G_{0}$, all the higher peaks would require the study of bigger
contacts with even more atoms in the central region.

It is important to remark that out of $50$ simulations we have only
observed the formation of three short chains with $3$, $4$ and $5$
atoms in each case. This is in strong contrast to the case of Au,
where chains were encountered much more frequently and with more chain
atoms.\cite{Dreher2005} Short atomic Ag chains of up to three atoms
have also been observed in experiments.\cite{AIYansonPhD,Rodrigues2002}

\begin{figure}
\includegraphics[  scale=0.25,
  keepaspectratio]{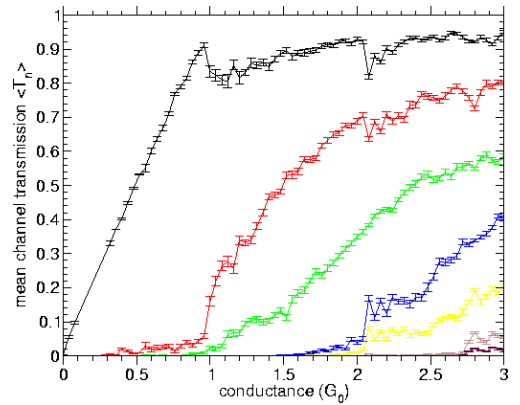}

\caption{\label{cap:Ag-avchannels}(Color online) Mean value of the transmission
coefficient $\left\langle T_{n}\right\rangle $ as a function of the
conductance for Ag ($4.2$ K, {[}$001${]} direction, $50$ contacts).
The error bars indicate the mean error.\cite{remarkmeanchannel}}
\end{figure}

Another important piece of information can be obtained from the analysis
of the mean channel transmission (averaged over all contacts) as a
function of the conductance, which is shown in Fig.~\ref{cap:Ag-avchannels}.\cite{remarkmeanchannel}%
 Here, one can see that the conductance region below $1\, G_{0}$
is largely dominated by a single channel. Above $1\, G_{0}$ a sharp
onset of the second transmission channel can be observed, the third
channel increasing more continuously. At $2\, G_{0}$ again an onset
of the fourth and fifth channel are visible.

These results can be related to the experimental observation on noble
metals made by Ludoph \textit{et al.},\cite{Ludoph1999,Ludoph2000}
namely the principle of the {}``saturation of channel transmission''.
This principle says that there is {}``a strong tendency for the channels
contributing to the conductance of atomic-size Au contacts to be fully
transmitting, with the exception of one, which carries the remaining
fractional conductance''.\cite{Ludoph1999} This tendency of the
channels to open one by one is evident for the first channel from
Fig.~\ref{cap:Ag-avchannels} and also experimentally the first peak
in the conductance histogram for Ag fulfills this principle best.\cite{Ludoph2000}
Concerning the higher conductances the finite size of our contacts
plays an increasingly restrictive role, but we are well in line with
the statement (made for Au, Ag and Cu) that {}``particularly the
second peaks in the histograms are also determined by other statistical
(probably atomic) properties of the contact''.\cite{Ludoph2000}

\section{Platinum atomic contacts\label{sec:Pt}}

Now, we turn to the analysis of Pt contacts. Pt is a transition metal
with $10$ valence electrons in the partially occupied $5d$ and $6s$
orbitals. The experiments reported so far show that in the case of
Pt the last conductance plateau lies typically above $1\, G_{0}$.
Consequently, the conductance histogram is dominated by the presence
of a broad peak centered around $1.5\, G_{0}$.\cite{AIYansonPhD,Smit2002}
Another remarkable feature of Pt contacts is the appearance of monoatomic
chains (with up to six atoms), which have a conductance ranging from
around $1.5$ to $1.0\, G_{0}$ as the length increases.\cite{Smit2001,Smit2003}
Moreover, complex oscillations of the conductance as a function of
the number of chain atoms are superimposed on top of such a decay.
Their origin has been explained in terms of a nearly half-filled $s$
band and the additional conduction channels provided by the almost
full $d$ bands.\cite{delaVega2004}

\subsection{Evolution of individual platinum contacts\label{sub:Pt-contact-evolution}}

\begin{figure}
\includegraphics[  scale=0.25]{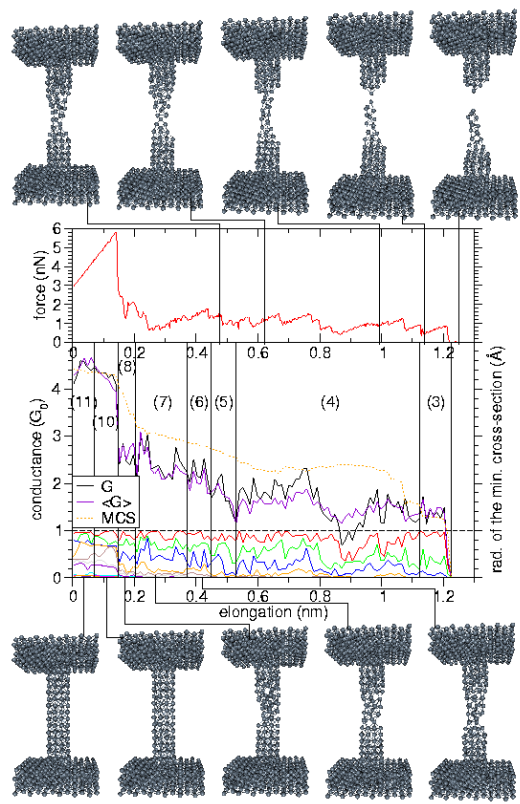}

\caption{\label{cap:Pt-contact-dimer}(Color online) Formation of a dimer
configuration for Pt ($4.2$ K, {[}$001${]} direction). The upper
panel shows the strain force as a function of the elongation of the
contact. In the lower panel the conductance $G$, the averaged conductance
$\left\langle G\right\rangle $, the MCS (minimum cross-section) radius
and the channel transmissions are displayed. Vertical lines separate
regions with different numbers of open channels ranging from $11$
to $3$.\cite{remarkchannelregions} Above and below these graphs
snapshots of the stretching process are shown.}
\end{figure}

In Fig.~\ref{cap:Pt-contact-dimer} a typical example of the formation
of a dimer configuration is shown. As before, beside the strain force,
we display the conductance, the averaged conductance, the MCS radius
and the channel transmissions.%
 The initial evolution is quite similar for all the $50$ Pt contacts
analyzed here. In this region, which corresponds to elongations below
$0.17$ nm, we observe between $11$ and $10$ open conduction channels.
After this region, and as in the case of Ag contacts, the conductance
evolves in a series of jumps which coincide with plastic deformations
(see the positions of break-ins in the sawtooth shape of the strain
force). However, in contrast to Ag, now we find strong conductance
fluctuations during the different elastic plateaus. The stretching
of the contact of Fig.~\ref{cap:Pt-contact-dimer} ends with the
formation of a dimer, which sustains three open channels and has a
conductance above $1\, G_{0}$ (see region with elongations between
$1.12$ nm and $1.22$ nm). This is again contrary to the Ag junctions
discussed above, where only a single dominant channel is observed
in the final stages before rupture.

On the other hand, the comparison between the conductance $G$ and
the averaged conductance $\left\langle G\right\rangle $ shows certain
deviations (see for instance the region with four channels). This
fact indicates that for Pt there is a much stronger variation of the
transmission as a function of the energy around the Fermi energy,
as compared with Ag. This is in agreement with the experimental finding
of nonlinear current-voltage characteristics for Pt as opposed to
linear ones for a noble metal like Au.\cite{Nielsen2002}

\begin{figure}
\includegraphics[  scale=0.25]{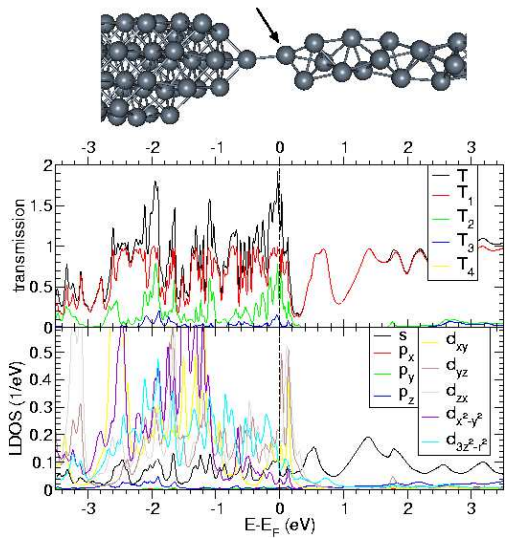}

\caption{\label{cap:Pt-T-LDOS-E-contact-dimer}(Color online) Pt contact of
Fig.~\ref{cap:Pt-contact-dimer} at an elongation of $1.18$ nm.
The total transmission $T$ is plotted as a function of the energy
together with the contributions from the different transmission channels
$T_{n}$. Additionally the LDOS (local density of states) is given
for an atom in the narrowest part of the contact, where the different
orbital contributions have been itemized. Above the figure the narrowest
part of the Pt contact is displayed in a magnified fashion and the
atom is indicated, for which the LDOS is shown.}
\end{figure}

The clear differences between the Pt and the Ag contacts can be traced
back to the difference in their electronic structure, as we now proceed
to illustrate. We show in Fig.~\ref{cap:Pt-T-LDOS-E-contact-dimer}
the LDOS for an atom in the narrowest part of the junction as a function
of the energy together with the transmission.%
 We have chosen a dimer configuration at an elongation of $1.18$
nm just before the rupture of the contact of Fig.~\ref{cap:Pt-contact-dimer}.
Notice the presence of a much more pronounced structure in the transmission
around the Fermi energy as compared to Fig.~\ref{cap:Ag-T-LDOS-E-contact2},
which can be attributed to the contribution of $d$ states. This fact
naturally explains the deviation between the conductance $G$ at $E_{F}$
and the averaged conductance $\left\langle G\right\rangle $ (cf.~Fig.~\ref{cap:Pt-contact-dimer}).
At the same time, the partially occupied $d$ orbitals are also responsible
for the larger number of open transmission channels (three in the
dimer region of Fig.~\ref{cap:Pt-contact-dimer}), as they provide
additional paths for electron transfer between the two electrodes.

From Fig.~\ref{cap:Pt-T-LDOS-E-contact-dimer} it is evident that
$d$ states play a major role for the conductance in Pt contacts.
The strong fluctuations of the conductance during the elastic stages
of stretching, as observed in Fig.~\ref{cap:Pt-contact-dimer}, point
out a high sensitivity of these $d$ states to the atomic configurations.
These two phenomena, namely the pronounced structure of the transmission
around the Fermi energy and the sensitivity of $d$ states to atomic
configurations are related. Indeed, a slight variation of $E_{F}$
for a fixed contact geometry has a similar effect on the conductance
as the modification of electronic level positions caused by a variable
contact geometry but a fixed Fermi energy. Ultimately, the sensitivity
of $d$ states to atomic configurations can be attributed to the spatial
anisotropy of the $d$ orbitals as compared to the spatially isotropic
$s$ orbitals, which are responsible for the conductance in Ag contacts.

\begin{figure}
\includegraphics[  scale=0.25]{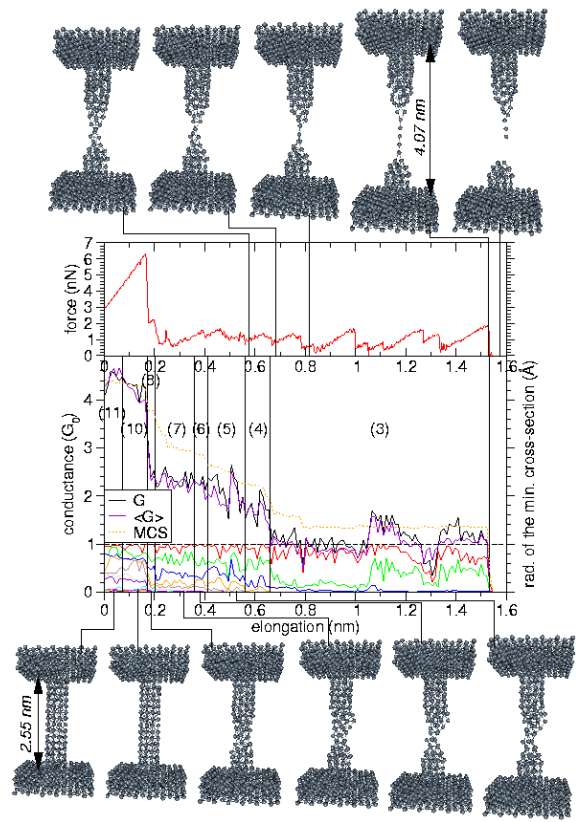}

\caption{\label{cap:Pt-contact-chain}(Color online) Formation of a five-atom
chain for Pt ($4.2$ K, {[}$001${]} direction). The upper panel shows
the strain force as a function of the elongation of the contact. In
the lower panel the conductance $G$, the averaged conductance $\left\langle G\right\rangle $,
the MCS (minimum cross-section) radius and the channel transmissions
are displayed. Vertical lines separate regions with different numbers
of open channels ranging from $11$ to $3$.\cite{remarkchannelregions}
Above and below these graphs snapshots of the stretching process are
shown.}
\end{figure}

Now we proceed to discuss the formation of chains in Pt contacts.
In the last stages of our simulations we often observe the formation
of special structures, namely linear chains of several atoms. In Fig.~\ref{cap:Pt-contact-chain}
we show the evolution of a Pt contact, which features a five-atom
chain before rupture.%
 As for the contact discussed previously, substantial fluctuations
in the conductance are visible even during the elastic stages, demonstrating
again the sensitivity of $d$ orbitals to atomic positions. The conductance
during the formation of the chain is mainly dominated by two channels,
but also a third one is contributing slightly. The first two channels
can be of nearly the same magnitude (see elongations above $1.1$
nm). After the dimer has formed, the transmission fluctuates around
$1\, G_{0}$. Compared with Ag, the conductance can, however, also
be higher than $1\, G_{0}$ due to the presence of a second and a
third open channel. The conductance of the last plateau is slightly
below the typical experimental value of $1.5\, G_{0}$,\cite{AIYansonPhD,Nielsen2003}
a fact that we shall discuss below.

During the formation of the chain (see three-channel region or elongations
above $0.8$ nm), the strain force exhibits a clear sawtooth behavior.
The abrupt jumps in the force after the long elastic stages signal
the incorporation of a new atom into the chain. Such incorporations
happen at elongations of $0.79$ nm (dimer), $1.00$ nm (three-atom
chain), $1.05$ nm (four-atom chain) and $1.27$ nm (five-atom chain).
Additional jumps at $0.83$ nm, $1.11$ nm and $1.33$ nm are due
to bond breakings at the chain ends. Note that the incorporation of
a new atom into an atomic chain does not always require long stretching
distances of the order of the nearest-neighbor distance. Because of
metastabilities depending on the geometry of the junction, they may
actually be much shorter, as can be inferred from the transition from
the three-atom chain to the four-atom chain.

\begin{figure}
\includegraphics[  scale=0.25]{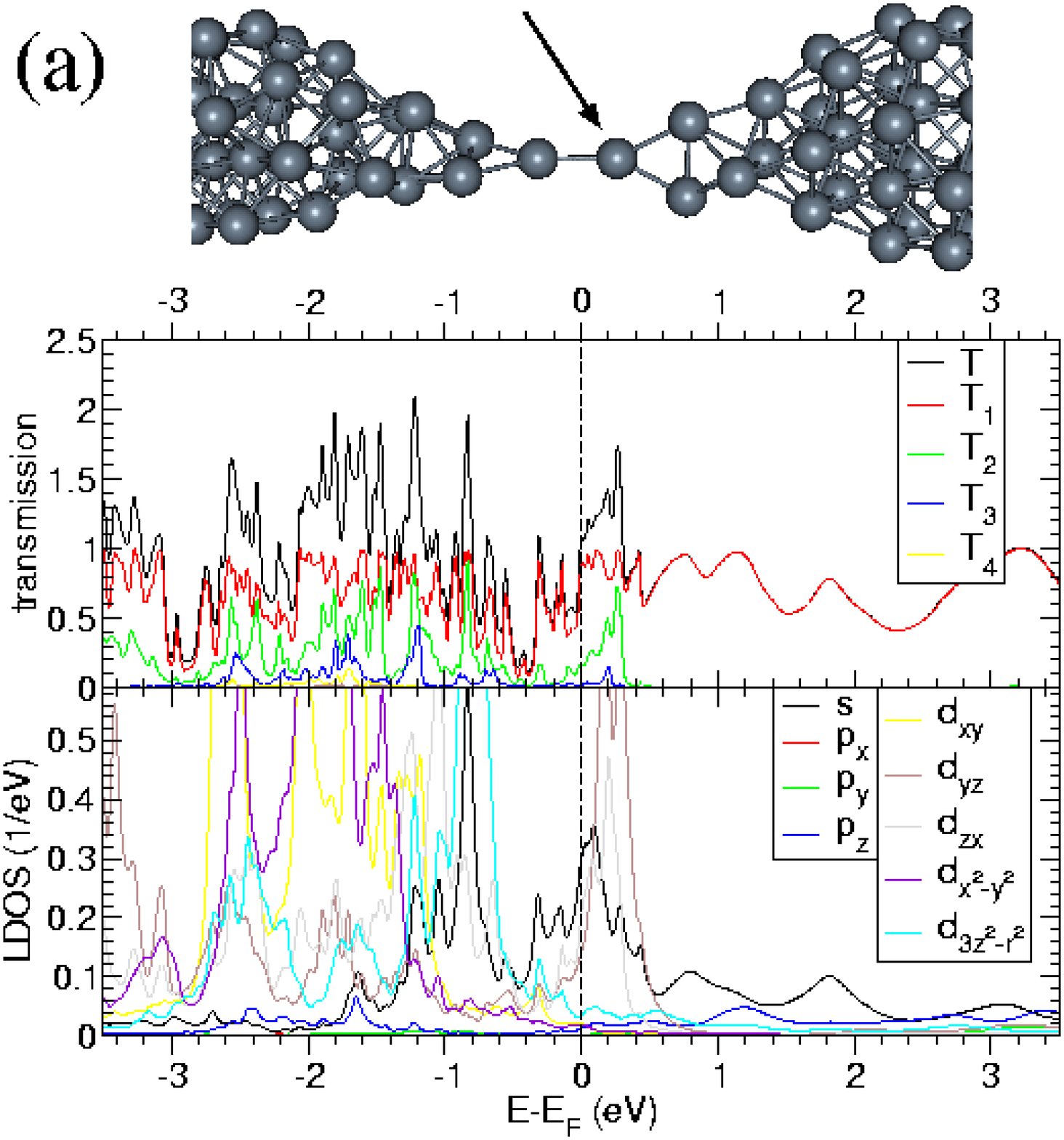}

\includegraphics[  scale=0.25]{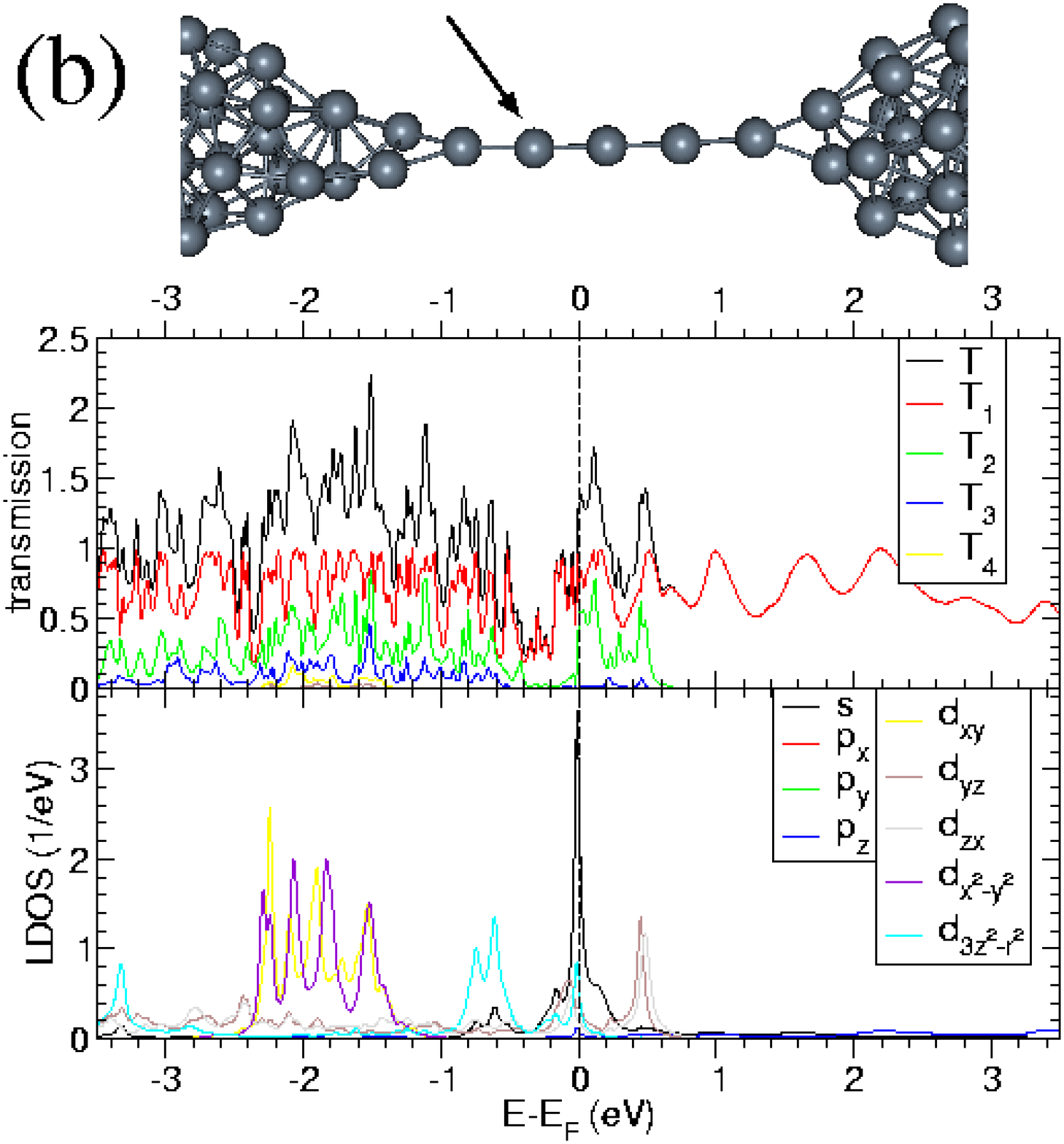}

\caption{\label{cap:Pt-T-LDOS-E-contact-chain}(Color online) Pt contact of
Fig.~\ref{cap:Pt-contact-chain} at an elongation of $0.95$ nm,
when the contact is forming a dimer (a) and at an elongation of $1.44$
nm, when the contact is forming a five-atom chain (b). In each case
the total transmission $T$ is plotted as a function of the energy
together with the contributions from the different transmission channels
$T_{n}$. Additionally the LDOS (local density of states) is given
for an atom in the narrowest part of the contact, where the different
orbital contributions have been itemized. Above each figure the narrowest
part of the Pt contact is displayed in a magnified fashion and the
atom is indicated, for which the LDOS is shown.}
\end{figure}

In order to explore changes in the electronic structure and their
influence on the transmission for the evolution from a dimer and to
long atomic chain, we analyze these two kinds of structures now in
more detail. In Fig.~\ref{cap:Pt-T-LDOS-E-contact-chain} we plot
the transmission and LDOS as a function of the energy, considering
as example the contact of Fig.~\ref{cap:Pt-contact-chain}.%
 As can be seen in Fig.~\ref{cap:Pt-T-LDOS-E-contact-chain}(a) for
the case of the dimer, the main contributions to the LDOS at the Fermi
energy come from the $s$, $d_{yz}$, $d_{zx}$ and $d_{3r^{2}-z^{2}}$
orbitals. Just like for the dimer structure investigated above, the
$d$ orbitals contribute significantly to the LDOS (cf.~Fig.~\ref{cap:Pt-T-LDOS-E-contact-dimer}).
While the energy dependence of the transmission looks qualitatively
similar, the LDOS changes dramatically when a long chain is formed
(see Fig.~\ref{cap:Pt-T-LDOS-E-contact-chain}(b)). We observe a
pinning of the $s$ and $d_{3r^{2}-z^{2}}$ states at the Fermi energy,
where the $s$ state is close to half filling corresponding to an
electronic $5d^{9}6s^{1}$ configuration of the Pt atom. (Notice also
the change in scale for the LDOS when going from Fig.~\ref{cap:Pt-T-LDOS-E-contact-chain}(a)
to Fig.~\ref{cap:Pt-T-LDOS-E-contact-chain}(b).) Comparing the energy
dependence of the transmission channels and the LDOS in Fig.~\ref{cap:Pt-T-LDOS-E-contact-chain}(b),
we can infer that the first channel is a linear combination of $s$,
$p_{z}$ and $d_{3r^{2}-z^{2}}$ orbitals ($l_{z}=0$), while the
second and third seem dominated by $d_{yz}$ and $d_{zx}$ orbital
contributions ($l_{z}=\pm 1$). These findings are perfectly in line
with Ref.~\onlinecite{delaVega2004}.

It is also noteworthy that when the $d$ states have decayed $1$
eV above the Fermi energy and the $s$ contribution dominates in the
LDOS, only a single channel is observed in the transmission for both
the dimer and the chain configuration (see Figs.~\ref{cap:Pt-T-LDOS-E-contact-dimer}
and \ref{cap:Pt-T-LDOS-E-contact-chain}). This would correspond exactly
to the situation described above for Ag wires, and demonstrates that
the differences between these two metallic contacts (Ag and Pt) are
mainly due to the different positions of their Fermi energy.

\subsection{Statistical analysis of platinum contacts\label{sub:Pt-statistical-analysis}}

\begin{figure}
\includegraphics[  scale=0.25]{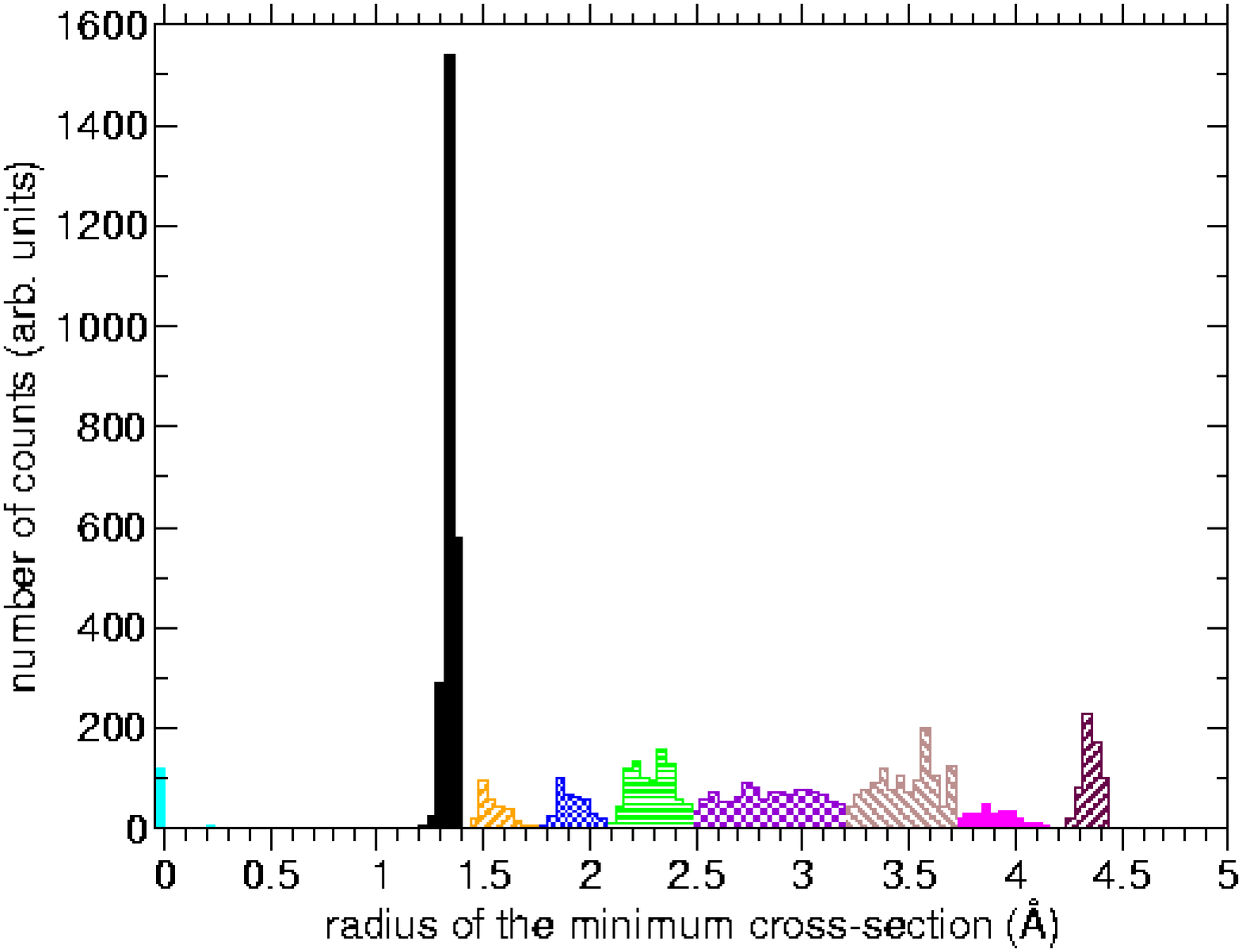}

\vspace{0.2cm}
\includegraphics[  scale=0.25]{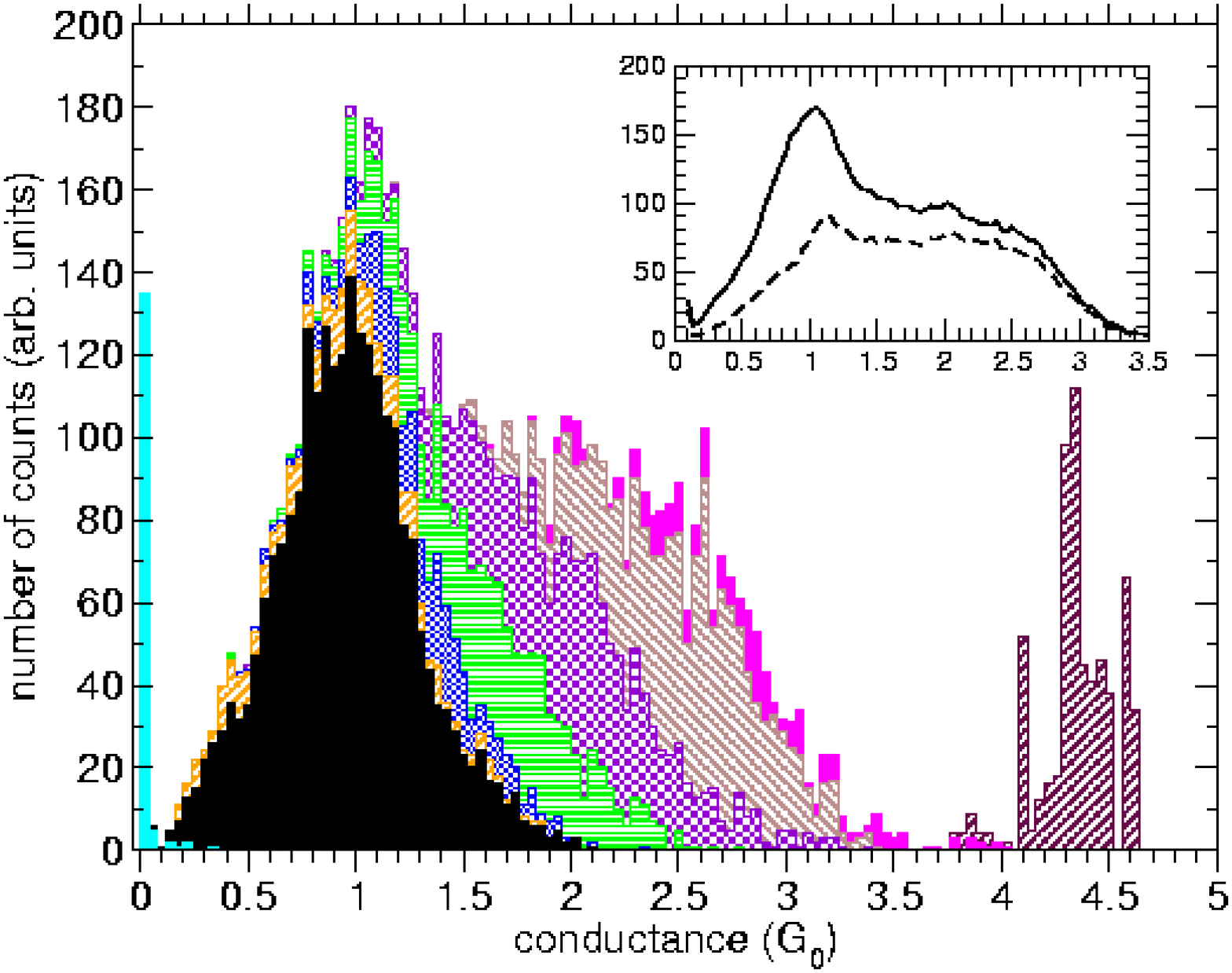}

\caption{\label{cap:Pt-mcs-condhisto}(Color online) MCS (minimum cross-section)
histogram (upper panel) and conductance histogram (lower panel) for
Pt ($4.2$ K, {[}$001${]} direction, $50$ contacts). In the MCS
histogram different regions of frequently occurring radii have been
defined with different pattern styles. The patterns in the conductance
histogram indicate the number of counts for conductances belonging
to the corresponding region of the MCS histogram. In the inset of
the lower panel the conductance histogram is displayed in the relevant
region in a smoothed version by averaging over six nearest-neighbor
points for all contacts (solid) and contacts with up to $8$ atoms
in the chain (dotted).}
\end{figure}

Putting together all the results for the $50$ Pt contacts simulated
in our study, we obtain the histograms for the MCS and conductance
shown in Fig.~\ref{cap:Pt-mcs-condhisto}.%
 The MCS histogram exhibits a very pronounced peak at radii corresponding
to dimer contacts and chains of atoms. Out of $50$ breaking events
we obtain $18$ chains, $17$ chains ranging from $5$ to $11$ atoms
and one with up to $19$ atoms. The tendency of Pt to form atomic
chains is consistent with experiments,\cite{AIYansonPhD,Smit2001}
but the ratio of chain formation is obviously higher than in the experiments.
This could partly be due to the thinness of the contacts that we investigate.
There exists experimental evidence for the formation of chains with
lengths up to six atoms,\cite{AIYansonPhD} while longer chains become
more and more unlikely. Therefore, our chains with more than eight
atoms seem somewhat artificial.

In the conductance histogram the low-lying MCS peak for dimers and
atomic chains gives rise to a very broad peak in the conductance histogram.
The position of this peak is centered around $1\, G_{0}$ rather than
$1.5\, G_{0}$, as in the experiment.\cite{AIYansonPhD,Nielsen2003}
If we exclude the longest chains (chains with more than eight atoms),
we obtain a conductance histogram with a very broad peak at $1.15\, G_{0}$
(cf.~inset in Fig.~\ref{cap:Pt-mcs-condhisto}).

Experimentally it has been shown that the peak at $1.5\, G_{0}$ shifts
to $1.8\, G_{0}$ for higher bias voltages.\cite{Nielsen2003} This
has been attributed to a structural transition, where atomic chains
are replaced by single-atom contacts. Thus, the conductance of dimers
and chains should be around $1.5\, G_{0}$ and the conductance of
single-atom contacts around $1.8\, G_{0}$. In Fig.~2 of Ref.~\onlinecite{Smit2003}
Smit \textit{et al}.~reported a decrease of the average conductance
from $1.5\, G_{0}$ to around $1\, G_{0}$ for increasing chain lengths.
This demonstrates that our broad distribution of conductances around
$1\, G_{0}$ in the conductance histogram (cf.~Fig.~\ref{cap:Pt-mcs-condhisto})
is not unreasonable, although the transmission for dimers and short
chains seems to be underestimated. A recent DFT study investigated
ideal Pt chains consisting of two to five atoms in the {[}$001${]}
direction.\cite{GarciaSuarez2005} Conductances between $2\, G_{0}$
and $1\, G_{0}$ were obtained with a trend toward $1\, G_{0}$ for
longer chains in agreement with experiment. The structure of the chains,
which in our case is linear, was zigzag-like. This could be another
explanation for the lower transmissions in our study.\cite{remarkPt-T1.5}

Although the peak position in the conductance histogram in Fig.~\ref{cap:Pt-mcs-condhisto}
is lower than in the experiments, we want to point out the strong
qualitative differences in comparison to Ag. While the first two MCS
peaks in the Ag histogram (cf.~Fig.~\ref{cap:Ag-mcs-condhisto})
are restricted to conductance values below $1\, G_{0}$, this is not
the case for Pt. Here, the first two peaks cover a range of conductance
values from as low as $0.1\, G_{0}$ up to $2\, G_{0}$. This is again
due to the contribution of the $d$ orbitals at the Fermi energy,
which leads to a higher number of open channels in the case of Pt,
as explained in Sec.~\ref{sub:Pt-contact-evolution}. Let us recall
that for Ag there is a single dominant transmission channel (and a
small second one), while for Pt there are usually three channels in
the last stages of breaking, and the second channel can be comparable
in magnitude to the first. As explained above, the extraordinary width
of the first peak in the conductance histogram for Pt can be attributed
to the sensitivity of $d$ states to the atomic configuration of the
contact.

\begin{figure}
\includegraphics[  scale=0.25,
  keepaspectratio]{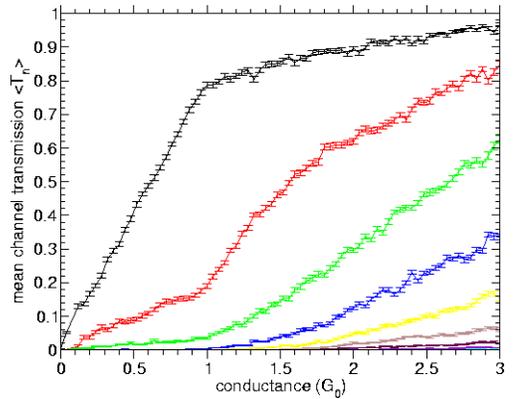}

\caption{\label{cap:Pt-avchannels}(Color online) Mean value of the transmission
coefficient $\left\langle T_{n}\right\rangle $ as a function of the
conductance for Pt ($4.2$ K, {[}$001${]} direction, $50$ contacts).
The error bars indicate the mean error.\cite{remarkmeanchannel}}
\end{figure}

This qualitative difference in the number of conduction channels is
illustrated in Fig.~\ref{cap:Pt-avchannels}, where we show the mean
value of the transmission coefficients as a function of the conductance.\cite{remarkmeanchannel}%
 Notice that as compared with the case of Ag (cf.~Fig.~\ref{cap:Ag-avchannels}),
there are contributions from the second and third channel already
present for conductances below $1\, G_{0}$. For conductances of $1.5\, G_{0}$
there are four or five channels on average.

In conclusion, the different behavior of Ag and Pt contacts stems
from the different electronic states present at the Fermi energy.
While for noble metals like Au and Ag it is located in the $s$ band,
its position is shifted downwards into the $d$ bands for Pt. Therefore,
in the latter case there are in general more open channels contributing
to the conductance. This confirms the statements of Scheer \textit{et
al.}\cite{Scheer1998} that the number of transmission channels is
determined by the chemical valence.

\section{Aluminum atomic contacts\label{sec:Al}}

Al is an example of the so-called $sp$-like metals. In the crystalline
form there are three valence electrons occupying partly the $3s$
and $3p$ bands around the Fermi energy. In this respect, Al has a
very different electronic structure as compared to Au, Ag or Pt, and
in this section we study how this electronic structure is reflected
in the conductance through Al atomic wires. Due to the technical problems
detailed below, this analysis will be considerably shorter than for
the other metals.

The experimental studies of the conductance of Al atomic-sized contacts
have shown several peculiar features.\cite{Krans1993,Scheer1997,Yanson1997,Cuevas1998b,AIYansonPhD}
For instance, Scheer \textit{et al.}\cite{Scheer1997}, making use
of the superconducting current-voltage characteristics to extract
the transmission coefficients, showed that usually three conduction
channels contribute to the transport, although the conductance of
the last plateau is typically below $1\, G_{0}$. This was explained
in Ref.~\onlinecite{Cuevas1998a} in terms of the contribution of
the $p$ orbitals to the transport. Exploiting conductance fluctuations,
the presence of several conduction channels for conductances above
$0.5\, G_{0}$ could subsequently be confirmed by another independent
experimental technique.\cite{Ludoph2000} As an additional peculiarity,
Al is one of the few multivalent metals which exhibits several pronounced
peaks in the conductance histograms at low temperatures.\cite{Yanson1997}
The first peak appears at around $0.8\, G_{0}$ and the next ones
at $1.9\, G_{0}$, $3.2\, G_{0}$ and $4.5\, G_{0}$. Furthermore,
the conductance plateaus in Al have a positive slope upon stretching,\cite{Krans1993,Scheer1997}
which is quite unique.

Again we simulated $50$ breaking events. Although we always observe
in the last stage of the nanocontacts either a single-atom contact
($36$ times), a dimer ($13$ times) and in one case a four-atom chain,
the single-atom contacts and dimers are often very short-lived configurations
and less stable than the corresponding Ag and Pt structures. We attribute
this to shortcomings in the semiempirical potential employed for Al
in this work. Previously it has been shown that this potential cannot
reproduce adequately the mechanical properties of an infinite Al chain.\cite{Albad2003}
This underestimation of the stability of thin wires is quite apparent
in our simulations, where the contacts break effectively at conductances
well above $1.5\, G_{0}$ and with several atoms present in the MCS.

\begin{figure}
\includegraphics[  scale=0.25]{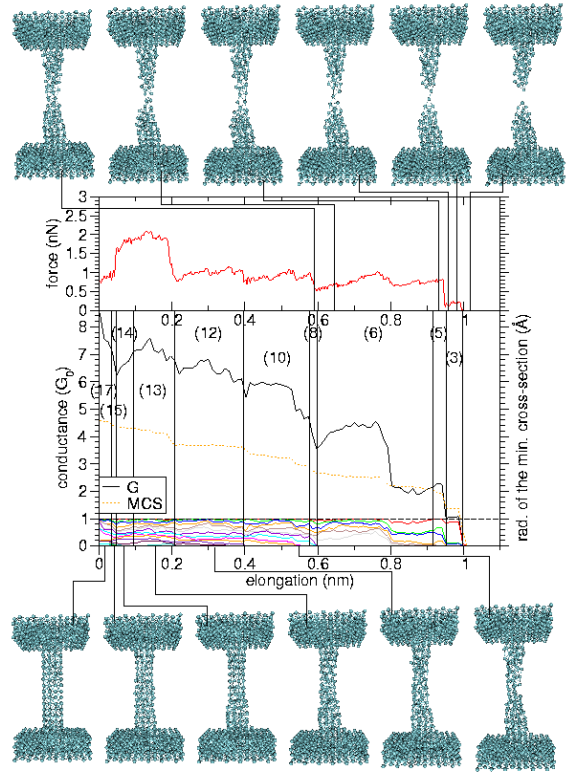}

\caption{\label{cap:Al-contact-good}(Color online) Formation of a dimer configuration
for Al ($4.2$ K, {[}$001${]} direction). The upper panel shows the
strain force as a function of the elongation of the contact. In the
lower panel the conductance $G$, the MCS (minimum cross-section)
radius and the channel transmissions are displayed. Vertical lines
separate regions with different numbers of open channels ranging from
$17$ to $3$.\cite{remarkchannelregions} Above and below these graphs
snapshots of the stretching process are shown.}
\end{figure}

This technical problem hindered the proper analysis of the statistical
properties of Al contacts. However, we could recover a few sensible
examples. One of the formations of a relatively stable dimer is displayed
in Fig.~\ref{cap:Al-contact-good}.%
 A region of three transmitting channels can be observed shortly before
contact rupture, and the conductance of the dimer configuration is
close to $1\, G_{0}$, which agrees nicely with the observations of
Scheer \textit{et al.}\cite{Scheer1997} The origin of these three
channels is, as explained in Ref.~\onlinecite{Cuevas1998a}, the contribution
of the partly occupied $sp$-hybridized valence orbitals of Al to
the transport. Before this region, a nice plateau around $2\, G_{0}$
is visible. Both features agree well with the peaks in the experimental
conductance histogram for Al close to $0.8\, G_{0}$ and $1.9\, G_{0}$.\cite{Yanson1997,AIYansonPhD}
More importantly, our results reproduce the peculiar positive slopes
of the last plateaus of the stretching curves, in compliance with
Refs.~\onlinecite{Krans1993,Scheer1997,Cuevas1998b,Jelinek2003}.

\section{Nickel atomic contacts\label{sec:Ni}}

During the last years a lot of attention has been devoted to the analysis
of contacts of magnetic materials.\cite{Costa1997,Ott1998,Ono1999,Viret2002,Elhoussine2002,Rodrigues2003,Untiedt2004}
(For a more complete list of references see Refs.~\onlinecite{Agrait2003,Untiedt2004}.)
In these nanowires the spin degeneracy is lifted, which can potentially
lead to interesting spin-related phenomena in the transport properties.
For instance, different groups have reported the observation of half-integer
conductance quantization either induced by a small magnetic field\cite{Ono1999}
or even in the absence of a field.\cite{Elhoussine2002,Rodrigues2003}
These observations are quite striking since such quantization requires
simultaneously the existence of a fully spin-polarized current and
perfectly open conduction channels.\cite{remarkhalfquantization}
With our present understanding of the conduction in these metallic
junctions, it is hard to believe that these criteria can be met, in
particular, in the ferromagnetic transition metals (Ni, Co and Fe).
As a matter of fact, in a more recent study by Untiedt \emph{et al.},\cite{Untiedt2004}
carried out at low temperatures and under cryogenic vacuum conditions,
the complete absence of quantization in atomic contacts of Ni, Co
and Fe has been reported, even in the presence of a magnetic field
as high as $5$ T. Several recent model calculations support these
findings.\cite{Martin2001,Delin2003,Bagrets2004,Jacob2005}

In this section we address the issue of the conductance quantization
and the spin polarization of the current with a thorough analysis
of Ni contacts. As described in Sec.~\ref{sub:Conductance-calculations}
we apply our method to a Hamiltonian with spin-dependent matrix elements.\cite{Haftel2004}

\subsection{Evolution of individual nickel contacts}

\begin{figure}
\includegraphics[  scale=0.25]{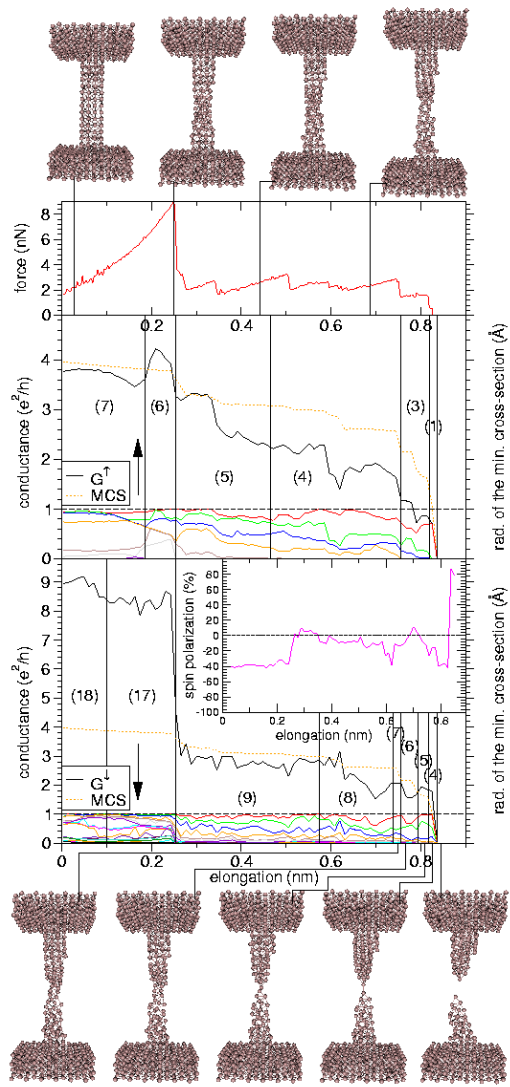}

\caption{\label{cap:Ni-contact1}(Color online) Formation of a dimer configuration
for Ni ($4.2$ K, {[}$001${]} direction). The upper panel shows the
strain force as a function of the elongation of the contact. In the
lower two panels the conductance $G^{\sigma }$, the MCS (minimum
cross-section) radius and the channel transmissions are displayed
for the respective spin component $\sigma $. Vertical lines separate
regions with different numbers of open channels ranging from $7$
to $1$ and $18$ to $4$, respectively.\cite{remarkchannelregions}
An inset shows the evolution of the spin polarization. Above and below
these graphs snapshots of the stretching process are shown.}
\end{figure}

In Fig.~\ref{cap:Ni-contact1} we show the evolution of the conductance
during the formation of a Ni dimer structure, which is the most common
geometry in the last stages of the breaking process.%
 Beside the evolution of the conductance and transmission eigenchannels
for both spin components separately, we have plotted the MCS radius,
strain force, spin polarization of the current and contact configurations.
The spin polarization $P$, shown in the inset of the lower panel,
is defined as \begin{equation}
P=\frac{G^{\uparrow }-G^{\downarrow }}{G^{\uparrow }+G^{\downarrow }}\cdot 100\%,\label{eq:spinpolcurrent}\end{equation}
 where $G^{\sigma }$ is the conductance of the spin component $\sigma $
(cf.~Eq.~(\ref{eq:conductanceG_spin})). Here, spin up ($\sigma =\uparrow $)
means majority spins and spin down ($\sigma =\downarrow $) minority
spins. Notice that in the last stages of the stretching the conductance
is dominated by a single channel for the majority spins, while for
the minority spin there are still up to four open channels. In the
final stages (see regions with three or one open channel(s) for $G^{\uparrow }$)
the conductance for the majority spin lies below $1.2e^{2}/h$, while
for the minority spin it is close to $2e^{2}/h$, adding up to a conductance
of around $1.2$-$1.6\, G_{0}$.

With respect to the evolution of the spin polarization of the current,
in the beginning of the stretching process it takes a value of around
$-40\%$, i.e.~the conductance of the minority-spin component outweighs
that of the majority-spin component. This is expected from the bulk
density of states of Ni. For this transition metal the Fermi level
lies in the $s$ band (close to the edge of the $d$ bands) for the
majority spins and in the $d$ bands for the minority spins. For this
reason, there is a larger number of conduction channels for minority-spin
component. This value of $P$ is indeed quite close to the value of
the spin polarization of the bulk density of states at the Fermi energy,
which in our model is equal to $-40.5\%$. As the contact geometry
starts changing, the spin polarization of the current begins to fluctuate.
It increases even to values of above $0\%$, but keeps a tendency
towards negative values, until it starts increasing to over $+80\%$
in the tunneling regime, when the contact is broken.

\begin{figure}
\includegraphics[  scale=0.25]{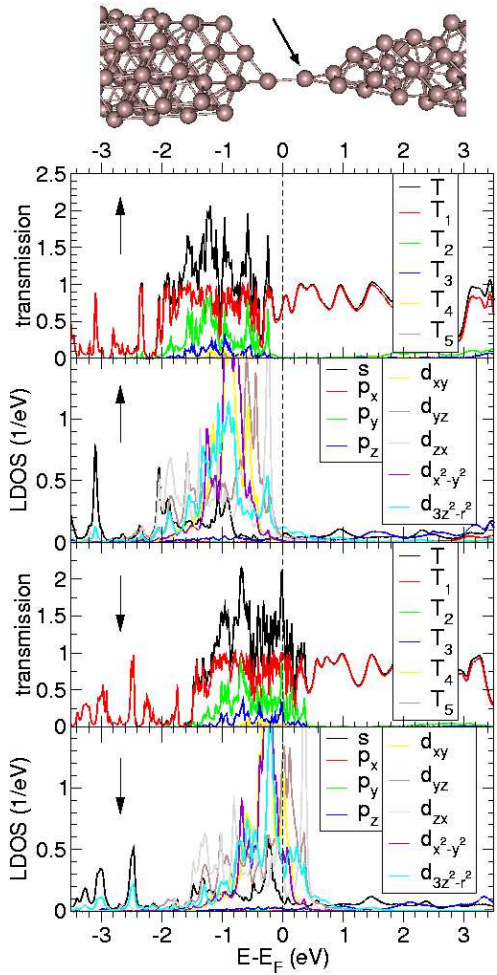}

\caption{\label{cap:Ni-T-LDOS-E-ud}(Color online) Ni contact of Fig.~\ref{cap:Ni-contact1}
at an elongation of $0.83$ nm. The transmission is plotted as a function
of the energy together with the contributions from the different transmission
channels $T_{n}^{\sigma }$ for the respective spin component $\sigma $.
Additionally the LDOS (local density of states) is given for each
spin component for an atom in the narrowest part of the contact, where
the different orbital contributions have been itemized. Above the
figure the narrowest part of the Ni contact is displayed in a magnified
fashion and the atom is indicated, for which the LDOS is shown. }
\end{figure}

Let us now try to gain further insight into these findings. We show
in Fig.~\ref{cap:Ni-T-LDOS-E-ud} the transmission as a function
of the energy together with the LDOS for an atom in the narrowest
part of the constriction portrayed in the upper part of the figure.%
 It can be observed that the Fermi energy, as in bulk, is located
just at the edge of the $d$ states for the majority-spin component,
while it is inside the $d$ states for the minority-spin component.
The majority-spin component therefore exhibits a single transmission
channel, behaving like a noble metal (cf.~results for Ag in Sec.~\ref{sec:Ag}),
while there are several open channels for the minority-spin component
like in the case of a transition metal (cf.~results for Pt in Sec.~\ref{sec:Pt}).

Concerning the spin polarization of the current, the large density
of states at $E_{F}$ for the minority-spin component usually gives
rise to a higher number of open channels for the minority-spin component
than for the majority-spin component, which in turn leads to a negative
spin polarization of the current. However, this argument is just qualitative,
because the actual transmission of the channels cannot simply be predicted
from the LDOS. The conductance depends also on the overlap of the
relevant orbitals and on non-local properties like the disorder in
the contact region. As a counter example, Fig.~\ref{cap:Ni-contact1}
shows that also intervals of positive spin polarization can be found,
although the density of states of the minority-spin component is usually
higher than for the majority-spin component. This is particularly
dramatic in the tunneling regime at the end of the breaking process,
where for instance in Fig.~\ref{cap:Ni-contact1} we see that a value
of $P=+80\%$ is reached. Such a reversal of the spin polarization
is due to the fact that the couplings between the $d$ orbitals of
the two Ni tips decrease much faster with distance than the corresponding
$s$ orbitals. As will be discuss further below, the result is typically
a reduction of the minority-spin conductance and therefore a positive
value of $P$.

We would like to point out that the contribution of the minority-spin
component to the conductance is very sensitive to changes in the configuration.
As is evident from Fig.~\ref{cap:Ni-contact1}, the minority spin
shows stronger fluctuations than the majority spin as a function of
the elongation. Again, this is a consequence of the fact that the
minority-spin contribution is dominated by the $d$ orbitals, which
are anisotropic and therefore more susceptible to disorder than the
$s$ states responsible for the conductance of the majority spins.
The sensitivity to atomic configurations is in agreement with the
findings for Ag and Pt as discussed above, where stronger fluctuations
of the conductance are seen for the transition metal Pt, as compared
with the noble metal Ag.

\subsection{Statistical analysis of nickel contacts}

\begin{figure}
\includegraphics[  scale=0.25,
  keepaspectratio]{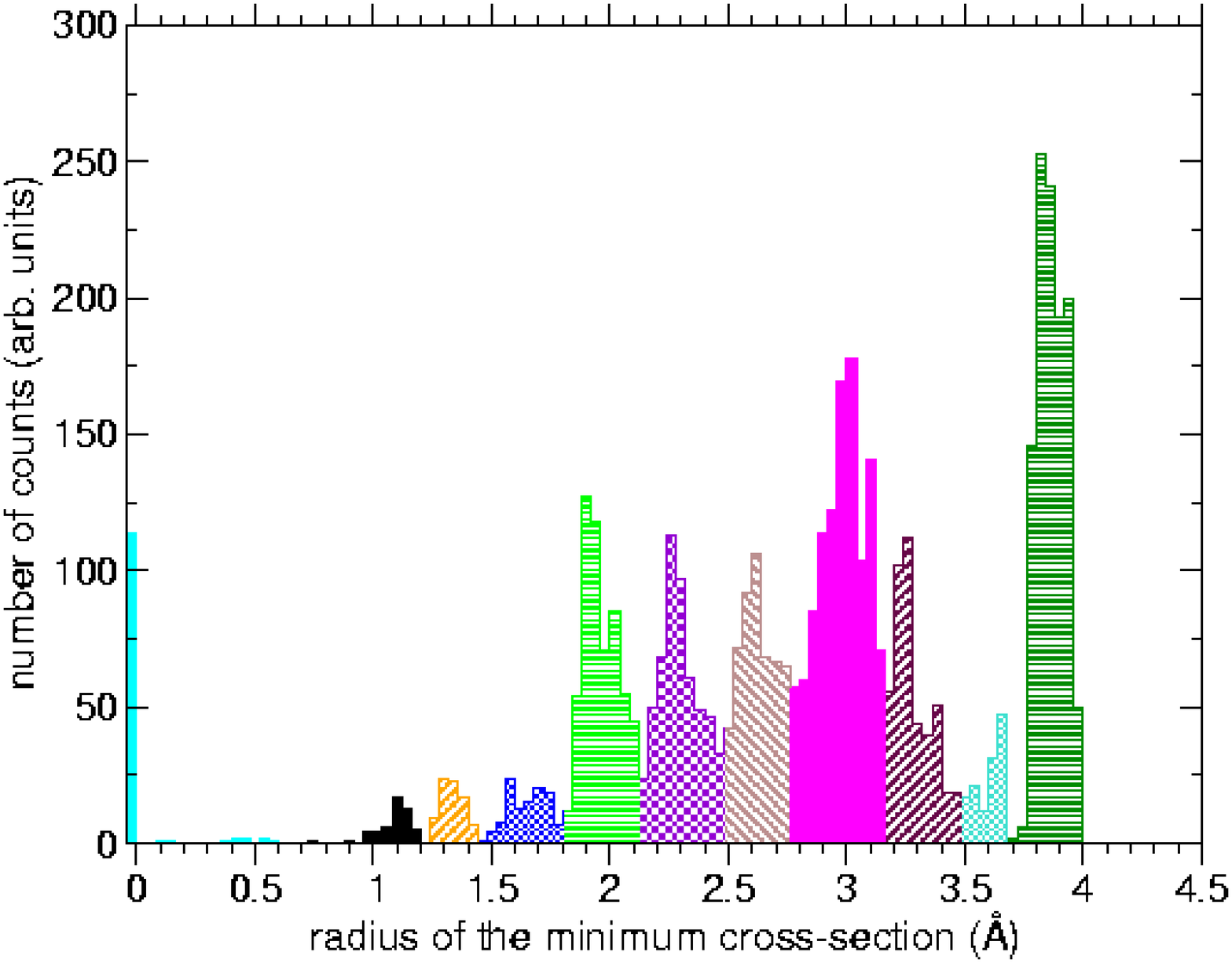}

\vspace{0.1cm}
\includegraphics[  scale=0.25,
  keepaspectratio]{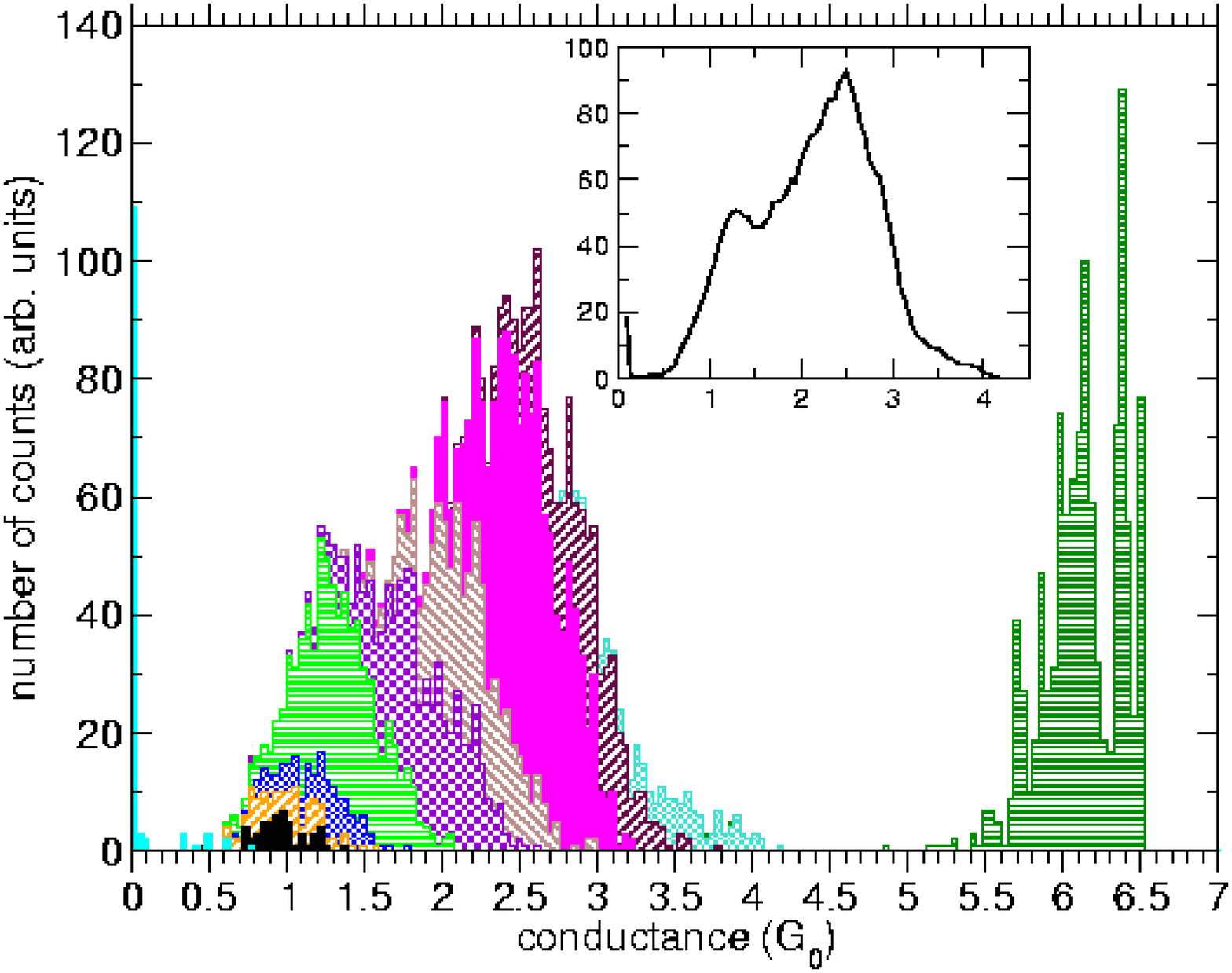}

\caption{\label{cap:Ni-mcs-condhisto}(Color online) MCS (minimum cross-section)
histogram (upper panel) and conductance histogram (lower panel) for
Ni ($4.2$ K, {[}$001${]} direction, $50$ contacts). In the MCS
histogram different regions of frequently occurring radii have been
defined with different pattern styles. The patterns in the conductance
histogram indicate the number of counts for conductances belonging
to the corresponding region of the MCS histogram. In the inset of
the lower panel the conductance histogram is displayed in the relevant
region in a smoothed version by averaging over six nearest-neighbor
points.}
\end{figure}

For the Ni contacts we did not observe the formation of any chain
in the $50$ simulated stretching processes. As a consequence, only
a small first peak is visible in the MCS histogram (see Fig.~\ref{cap:Ni-mcs-condhisto}).%
 This peak originates from the dimer configurations, which usually
form before the contacts break. In the conductance histogram there
is a shoulder at around $1.3\, G_{0}$. Part of this first peak is
buried under the subsequent conductance peak with its maximum at $2.5\, G_{0}$.
This second very broad peak is mainly influenced by the starting configuration,
which means that the small size of our contacts might hide part of
the peak structure in the conductance histogram. According to the
MCS regions contributing to the shoulder in the Ni conductance histogram,
the first peak is mainly composed of thick contacts (MCS of around
$2$ $\textrm{Å}$). This also explains the large broadening of the
histogram peak, since for thick contacts, there is more configurational
variability.

Concerning the comparison with measurements, the shoulder at $1.3\, G_{0}$
in our results is in agreement with the experimental conductance histogram,
where a particularly broad peak between $1.1\, G_{0}$ and $1.6\, G_{0}$
is observed.\cite{Untiedt2004} Our calculations indicate that this
peak contains contributions from high-MCS regions. The remarkable
width of the first peak in the experimental conductance histogram
is then explained by the configurational variability of thick contacts
in conjunction with the contribution of configurationally sensitive
$d$ states to the conductance of the minority-spin component. However,
this interpretation requires further discussion. Usually the first
peak in the experimental conductance histograms is believed to arise
from single-atom contacts and dimers.\cite{Scheer1998} With respect
to the problems encountered for Al (cf.~Sec.~\ref{sec:Al}), it
may be that the employed EMT potential for Ni underestimates the stability
of single-atom and dimer configurations in a similar manner. As a
consequence the contribution of such configurations to the first peak
in the conductance histogram may be underestimated in our calculations.
In addition, as mentioned above, this first peak in the conductance
histogram is not well separated from contributions with a high MCS,
which are influenced by our starting configuration. Simulations of
thicker contacts and more sophisticated calculations of the contact
geometry may be needed to clarify the robustness of our findings.

\begin{figure}
\includegraphics[  scale=0.22,
  keepaspectratio]{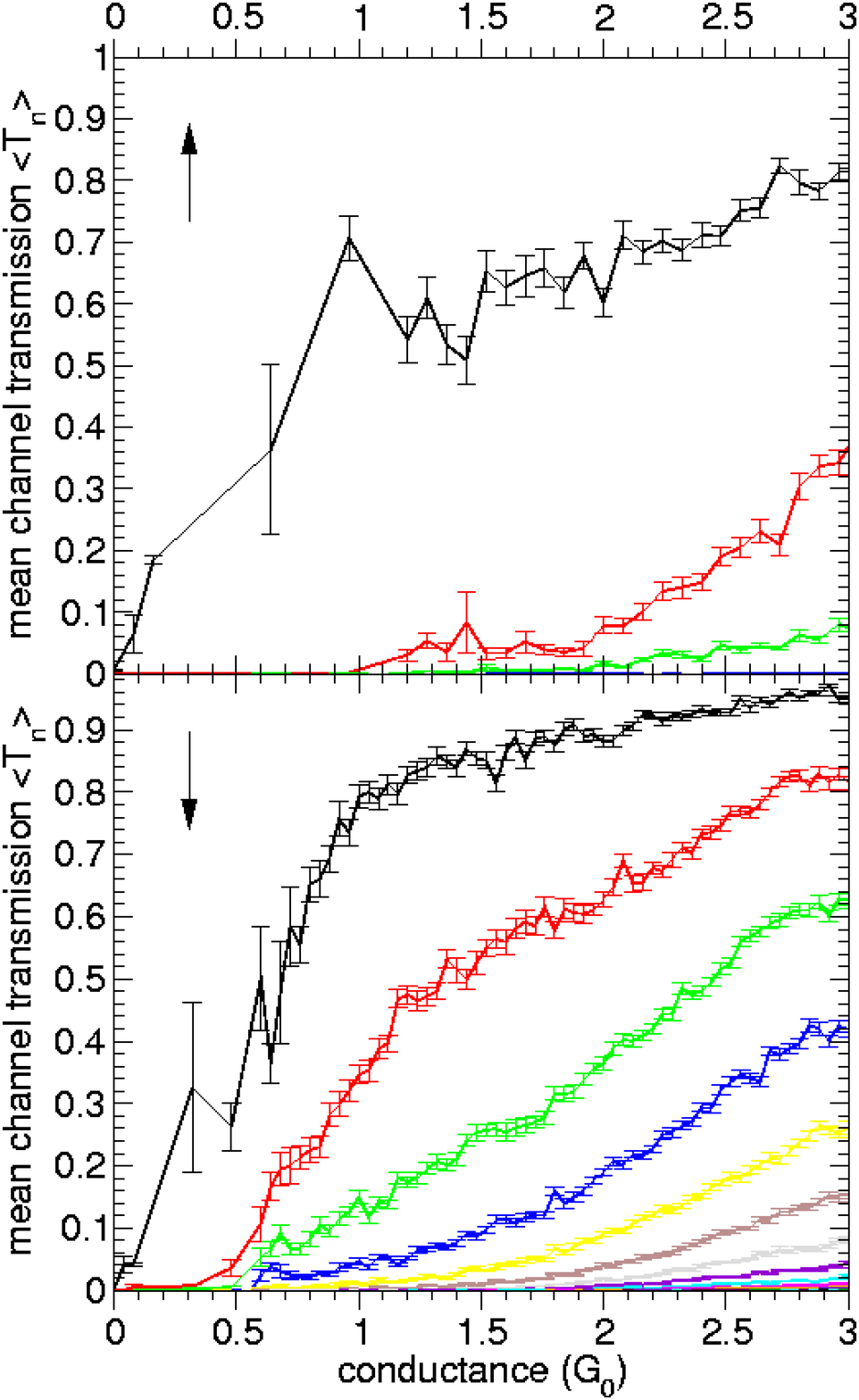}

\caption{\label{cap:Ni-avchannels-ud}(Color online) Mean value of the transmission
coefficient $\left\langle T_{n}^{\sigma }\right\rangle $ for the
respective spin component $\sigma $ as a function of the conductance
for Ni ($4.2$ K, {[}$001${]} direction, $50$ contacts). The error
bars indicate the mean error.\cite{remarkmeanchannel}}
\end{figure}

Regarding to the mean channel transmission of the spin-components
as a function of the conductance,\cite{remarkmeanchannel} the minority-spin
component exhibits more transmission channels than the majority-spin
component (see Fig.~\ref{cap:Ni-avchannels-ud}).%
 This further illustrates our previous argument, where we explained
that the majority-spin component possesses an Ag-like character, while
the minority-spin component behaves more Pt-like. Note also that the
first channel for the majority-spin component opens up remarkably
slowly compared with Ag (cf.~Fig.~\ref{cap:Ag-avchannels}).

\begin{figure}
\includegraphics[  scale=0.25,
  keepaspectratio]{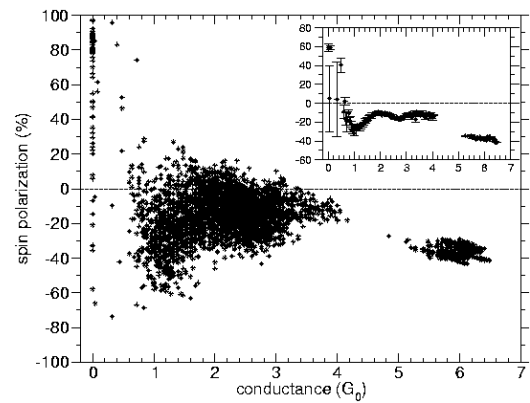}

\caption{\label{cap:Ni-spinpolarization}Spin polarization of the current
as a function of the conductance. All the data points for the spin
polarization are plotted in the graph, while in the inset their arithmetic
mean and the corresponding mean error are displayed.\cite{remarkspinpolinset}}
\end{figure}

Now we want to address the question of how the spin polarization of
the current is influenced by configurational changes. For this purpose,
we show in Fig.~\ref{cap:Ni-spinpolarization} the spin polarization
$P$ as a function of the conductance for all the $50$ simulated
breaking events. %
 As it could already be observed in the simulation of a single breaking
event (cf.~Fig.~\ref{cap:Ni-contact1}), the spin polarization of
all the contacts starts at a value of $-40\%$, when the contact is
close to its starting configuration. As explained above, this value
for the spin polarization of the current coincides rather well with
the polarization of the bulk density of states at the Fermi energy.
As the contact is stretched, also the diversity of geometrical configurations
increases and the spin polarization values are widely spread, ranging
from around $-60\%$ to $20\%$. There is a tendency towards negative
spin polarizations, as can be observed in the inset of Fig.~\ref{cap:Ni-spinpolarization}.
The average spin polarization varies between $-30\%$ and $-10\%$
for conductances above $0.6\, G_{0}$. As described in the previous
subsection, these variations arise from the high sensitivity of the
minority-spin conductance to atomic positions, as compared to the
less sensitive majority-spin conductance. The trend towards negative
$P$ values can be explained by the higher number of states present
at the Fermi energy for the minority-spin component as opposed to
the majority-spin component. 

In the region of conductances below $0.6\, G_{0}$ the number of points
is comparatively lower, which explains the partly bigger error bars.
Nevertheless, the number of realizations is still enough to see the
spreading of $P$ values over an even wider interval than in the contact
regime, together with an average tendency towards positive values.
We attribute this trend of reversed spin polarizations to the faster
radial decay of the hoppings between the $d$ orbitals that dominate
the minority-spin contribution to the conductance, as compared with
the $s$ orbitals that dominate majority-spin contribution. The faster
decay with distance overcomes in the tunneling regime the effect of
the higher density of states of the $d$ bands versus the $s$ bands.

\section{Mechanical properties of metallic atomic contacts\label{sec:Discussion}}

Experimentally it is possible to measure simultaneously the conductance
and the strain force during the breaking of nanowires.\cite{Agrait1995}
Special attention has been devoted to the force in the very last stage
of the stretching process.\cite{RubioB2001} For this reason, we present
in this section a detailed analysis of this breaking force for the
different metals discussed above, including the results for Au of
Ref.~\onlinecite{Dreher2005}. In addition, the exotic atomic chain
structures will be investigated further.

\begin{figure}
\includegraphics[  scale=0.25]{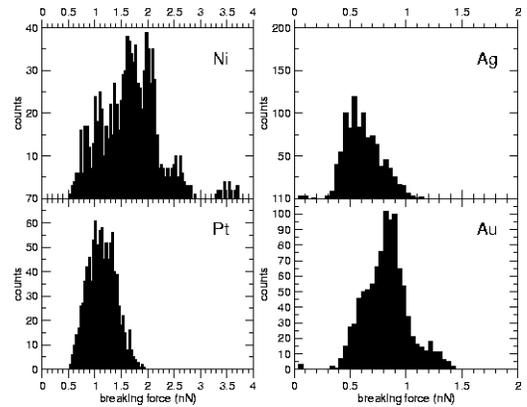}

\caption{\label{cap:bfHisto}Histogram of the force needed to break atomic
contacts of the metals Ag, Au, Pt and Ni. For every contact the highest
$20$ force values of the last $30$ recorded geometries before the
point of rupture are gathered. The force data for $50$ contacts of
the respective metal are assembled in the respective histograms.}
\end{figure}

Using the $50$ contacts that we have simulated for the different
metals, we construct histograms of the breaking force in the following
way: We consider the last $30$ recorded atomic configurations before
the point of rupture of the contact. Out of them the $20$ highest
values of the strain force are assembled in a force histogram, combining
the data from all $50$ contacts.\cite{remarkbfhisto} The breaking
force histograms obtained for the different metals are shown in Fig.~\ref{cap:bfHisto}.%
 For all elements, except for Ni, a clear maximum is visible in the
center of a broad distribution of force values. We will address later,
why Ni forms an exception in our simulations.

It is elucidating to compare the values of the breaking force obtained
in the simulations with the corresponding forces in bulk. For this
purpose, we use the {}``universal'' binding energy function, suggested
in Ref.~\onlinecite{Rose1998}, to get a rough estimate for the breaking
force expected for a bulk bond (see Appendix \ref{sec:Bulk-force}
for details).

Values for the breaking forces are put together in Tab.~\ref{cap:bfTab}.%
\begin{table}
\begin{tabular}{|c|c|c|c|c|}
\hline 
metal&
Ag&
Au&
Pt&
Ni\\
\hline
\hline 
EMT&
$0.40$-$0.75$&
$0.55$-$1.00$&
$0.80$-$1.45$&
$1.00$-$2.15$\\
\hline 
bulk&
$0.57$&
$0.85$&
$1.31$&
$0.89$\\
\hline
\end{tabular}

\caption{\label{cap:bfTab}Breaking forces in nN for the metals Ag, Au, Pt
and Ni. The values in the column called {}``EMT'' (effective medium
theory) are read off from the force histograms in Fig.~\ref{cap:bfHisto}
and {}``bulk'' refers to Eq.~(\ref{eq:bulk_force_final}).}
\end{table}
 The expression for the breaking force in Eq.~(\ref{eq:bulk_force_final})
needs to be considered as a rough estimate of the force needed to
break a bulk-like bond. Concerning a comparison of this bulk estimate
and the EMT results, it needs to be recalled that the EMT employed
in the MD simulations considers by construction the experimentally
verified increase of atomic bonding energies for low coordination.\cite{Jacobsen1996}
Breaking forces for low-coordinated chains have been shown to be two
to three times larger than bulk-like bonds and bond breaking may take
place at distances well before the inflection point of the bulk estimate
(cf.~Appendix \ref{sec:Bulk-force}).\cite{Bahn2001} Another difference
is that the forces listed under {}``bulk'' are estimates for breaking
forces of a single bond. This is not necessarily the case for the
result called {}``EMT''. The EMT results are based on the stretching
of the nanocontacts in our MD simulations. If the contact breaks while
more than one atom resides in the MCS, several atomic bonds might
be contributing to the breaking force of the contacts. This implies
that the resulting force could be higher than the breaking force for
a single bond.

For elements with a large peak in the MCS histogram at single-atom
radii, like for the elements Au and Pt, which form chains, usually
the contacts break after the formation of a dimer or atomic chain.
As a consequence, for Ag, Au and Pt single atomic bonds are probed
in the EMT results. For all these elements, the force estimated from
bulk considerations agrees surprisingly well with the EMT results.
For Ni however, there is a discrepancy between the breaking force
determined with EMT and the bulk prediction. We attribute this to
the fact that its MCS histogram does not display a pronounced peak
for dimer structures (cf.~Fig.~\ref{cap:Ni-mcs-condhisto}), indicating
that Ni dimers are less stable than dimers of the other investigated
metals (Ag, Au and Pt). On account of this the breaking force typically
contains contributions from more than a single atomic bond, and is
therefore higher than the force of the bulk estimate. The contributions
of several bonds also explain the broad distribution without a clear
maximum for Ni in Fig.~\ref{cap:bfHisto}.

The absolute values of our breaking forces in Tab.~\ref{cap:bfTab}
need not be quantitative, as the investigations of Rubio-Bollinger
\textit{et al.}\cite{RubioB2001} show. While our EMT-breaking force
for Au coincides well with their value of {}``around $1$ nN'',
they found that DFT calculations are in better agreement with the
experimentally measured breaking force of $1.5$ nN.

Coinciding with the DFT-simulations by Bahn \textit{et al}.\cite{Bahn2001}
the ordering of breaking force strengths for the different metals
as predicted by the bulk estimate is $F_{\textrm{Ag}}<F_{\textrm{Au}}<F_{\textrm{Ni}}<F_{\textrm{Pt}}$,
where $F_{\textrm{x}}$ is the breaking force for the material x.
The EMT results modify this ordering slightly by interchanging Pt
and Ni.

Before we conclude, we want to investigate the appearance and structural
properties of the peculiar atomic chain structures in more detail.
The general mechanism behind the chain formation during a stretching
process is an increase in bond strength between low-coordinated atoms.\cite{BahnPhd2001,Bahn2001,RubioB2001,Dreher2005}
Independent of the metal under investigation, we observed that contacts
in which an atomic chain has formed always break because of a bond
rupture at the chain ends. The higher bond strength for low-coordinated
atoms explains this phenomenon. Namely, the terminal atoms in the
chain are connected with the thicker part of the contact, and possesses
a higher coordination number than the other chain atoms. As a consequence
the bonds at the chain ends are weaker than the bonds in the interior
of the chain.\cite{BahnPhd2001}

\begin{figure}
\includegraphics[  scale=0.27]{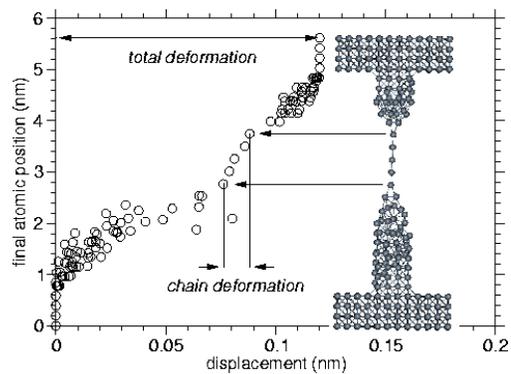}

\caption{\label{cap:Pt-chain-displacement}(Color online) The atomic displacements
for the Pt contact of Fig.~\ref{cap:Pt-contact-chain} are shown
in the last elastic stage before rupture (change in coordinates between
initial and final elongations of $L_{i}=1.37$ nm and $L_{f}=1.49$
nm). On the abscissa the displacement of each atom is plotted, while
on the ordinate the positions of the atoms can be seen at the end
of the elastic stage (elongation $L_{f}$). To the right the final
configuration is displayed. The atomic displacements and positions
have both been projected onto the stretching direction ($z$ axis).}
\end{figure}

We want to illustrate the mechanical properties of an atomic chain
considering as example the Pt contact of Fig.~\ref{cap:Pt-contact-chain}.
In Fig.~\ref{cap:Pt-chain-displacement} we plot the atomic displacements
for this Pt contact projected onto the stretching direction ($z$
axis) in the final elastic stage for elongations of $L_{i}=1.37$
nm and $L_{f}=1.49$ nm.%
 The $z$-projected displacement is defined as $d_{z,j}=R_{z,j}(L_{f})-R_{z,j}(L_{i})$,
where $R_{z,j}$ is the $z$-component of atom $j$, and $L_{f}$
($L_{i}$) is the final (initial) elongation. (Additionally we add
an offset to $d_{z,j}$, such that the fourth layer in the lower electrode
has zero displacement.) Due to the low coordination of the chain atoms
and the associated higher bond strength as compared to interatomic
bonds of the other atoms in the central part of the nanowire, the
chain is expected to be particularly stable. For this reason the chain
atoms should stay close to each other in a displacement plot during
an elastic stage of stretching. Instead, most of the displacement
should take place in the regions of more highly coordinated atoms
in the central part of the nanowire. Exactly this is visible in Fig.~\ref{cap:Pt-chain-displacement}.
Note that a similar analysis has been performed by Rubio-Bollinger
\textit{et al}.\cite{RubioB2001} for a Au chain.

Finally, we want to comment on experimental results of Ref.~\onlinecite{Smit2001}.
There, Smit \textit{et al.}~compare the tendency of formation of
atomic chains for the neighboring $4d$ and $5d$ elements, namely
Rh, Pd and Ag compared to Ir, Pt and Au. They find a higher occurrence
of chains for the $5d$ elements as compared to $4d$ elements, and
explain this by a competition between $s$ and $d$ bonding. From
their data\cite{remarkSmitAgAu} we extract an enhancement factor
of chain formation of $3.28$ for Au compared to Ag. Taking the ratio
between the content of the first MCS peak in the histograms, which
corresponds to dimers and atomic chains, for Ag and Au normalized
by the complete area of the MCS histograms (cf.~Fig.~\ref{cap:Ag-mcs-condhisto}
and Fig.~9 of Ref.~\onlinecite{Dreher2005}), we obtain a value
of $3.09$, in good agreement with their experiments. Bahn \textit{et
al}.\cite{Bahn2001} pointed out that the chain formation depends
sensitively on the initial atomic configuration. In general we believe
that chain formation in our thin geometries might be enhanced compared
to experimental conditions. Nevertheless the chain enhancement factor,
as it is a relative measure, might be robust.

\section{Conclusions\label{sec:Conclusions}}

In summary, we have analyzed the mechanical and electrical properties
of Ag, Pt and Ni nanojunctions. Using a combination of classical MD
simulations and transport calculations based on a TB model supplemented
with a local charge neutrality condition, we have studied the origin
of the experimentally observed characteristic features in the conductance
histograms of these metals. The ensemble of our results indicates
that the peak structure of the low-temperature conductance histograms
originates from an artful interplay between the mechanical properties
and the electronic structure of the atomic-sized contacts. 

We have found strong qualitative differences between these metals.
In the case of Ag wires, we observe a first peak at $1\, G_{0}$ in
the conductance histogram, resulting from single-atom contacts and
dimers in good agreement with experiments.\cite{AIYansonPhD} In the
last stages of the stretching process the transport is dominated by
a single conduction channel, which arises mainly from the contribution
of the $5s$ orbitals. We find practically no formation of monoatomic
chains, as opposed to Au wires.\cite{Dreher2005} To be precise, the
chain formation is found to be suppressed by a factor of three compared
to Au, which is again consistent with the experimental observations.\cite{Smit2001}

In the case of Pt contacts, the first peak in the conductance histogram
is mainly due to single-atom contacts and long atomic chains. However,
it also contains some contributions from contacts with larger MCS
radii. This peak is rather broad and centered around $1.15\, G_{0}$,
which is somewhat below the experimental value of $1.5\, G_{0}$.\cite{AIYansonPhD,Smit2002}
The differences in width and value of this conductance peak, as compared
with Ag, can be attributed to the key contribution of the $5d$ orbitals
to the transport. First, the $d$ orbitals provide additional conduction
channels. Commonly there are three open transmission channels in the
final stages of the Pt contacts. Second, these additional channels
naturally give rise to higher conductance values. Third, caused by
their spatial anisotropy the $d$ orbitals are much more sensitive
to changes in the contact geometry, which results in a larger width
of the histogram features.

With respect to Al the statistical analysis of the contacts was hindered
due to shortcomings in the employed EMT potential. However, for a
sensible example a region of three transmitting channels is observed
shortly before contact rupture, and the conductance of the dimer configuration
is close to $1\, G_{0}$, which agrees nicely with experimental observations\textit{.}\cite{Scheer1997}
These three channels originate from the contribution of the partly
occupied $sp$-hybridized valence orbitals of Al to the transport.
In addition, our results reproduce the peculiar positive slopes of
the last plateaus of the stretching curves.\cite{Krans1993,Scheer1997,Cuevas1998a,Jelinek2003}

In the case of ferromagnetic Ni, we have shown that the contacts behave
as a mixture of a noble metal (like Ag) and a transition metal (like
Pt). While the $4s$ orbitals play the main role for the transport
of the majority-spin electrons, the conduction of the minority-spin
electrons is controlled by the partially occupied $3d$ orbitals.
This follows from the position of the Fermi energy, which lies in
the $s$ band for the majority spins and in the $d$ bands for the
minority spins. In the conductance histogram we obtain a shoulder
at $1.3\, G_{0}$, whose large width can again be attributed to the
extreme sensitivity of the $d$ orbitals to atomic configurations.
On the other hand, we find that the spin polarization of the current
in the Ni contacts is generally negative, increasing and fluctuating
as the contacts narrow down and become disordered. In particular,
large positive values are possible in the tunneling regime, right
after the rupture of the wires. Once more, this behavior can be traced
back to the fact that the $d$ orbitals play a key role in the conductance
of the minority-spin component.

The mechanical properties of our nanocontacts have been analyzed in
detail with respect to breaking forces and the peculiar atomic chains.
Concerning the breaking forces a simple estimate for the maximal force
per bulk bond matches well the simulation results for Ag and Pt. However,
Ni shows deviations from the bulk estimate and an extraordinarily
broad distribution of breaking force values, which we attribute to
the generally larger thickness of the contacts at the breaking point,
meaning that the breaking force contains contributions of several
atomic bonds. Contacts with an atomic chain configuration were observed
to always tear apart due to a bond rupture at the chain ends in agreement
with previous simulations.\cite{BahnPhd2001} Pt atomic chains were
illustrated to exhibit an enhanced stability as compared to the remaining
atoms in thicker parts of the nanowire. 

Another important observation is that, although we obtain for every
metal a sequence of peaks in the MCS (minimum cross-section) histogram,
these peaks are smeared out in the conductance histograms. This indicates
that not only the narrowest part of the constriction determines the
conductance, but also the atomic configuration close to the narrowest
part plays a role. This finding challenges the direct translation
of peak positions in the conductance histogram into contact radii
via the Sharvin formula. However, we should also point out the limitations
of our modeling, in particular the small number of atoms present in
the junctions. Moreover, let us remind that we have focused our analysis
on low temperatures ($4.2$ K), where the atoms do not have enough
kinetic energy to explore the low-energy configurations. Both the
small number of atoms and the low temperature may cause an enhanced
atomic disorder of the contacts in the stretching process.

The effects of higher temperatures, different crystallographic orientations
of the contacts, other protocols of the stretching process with different
annealing, heating and relaxation times have not been addressed in
this study. A first-principles description of thick contacts, in which
both the mechanics and the electronic structure of the contacts are
treated at a higher level of accuracy and on an equal footing, should
be a major goal for the theory in the future. Experiments in which,
simultaneously to the recording of a conductance histogram, also the
contact geometries are observed, could help to validate the correlation
between conductance peaks and stable wire radii.

\section{Acknowledgments}

We thank E.~Scheer, Gerd Sch\"{o}n and R.H.M.~Smit for helpful
discussions. FP, JKV, MH and JCC were supported financially by the
Landesstiftung Baden-W\"{u}rttemberg within the {}``Kompetenznetz
Funktionelle Nanostrukturen'', the Helmholtz Gemeinschaft within
the {}``Nachwuchsgruppen-Programm'' (Contract No.~VH-NG-029) and
the DFG within the CFN. The granting of computer time at the Institut
für Nanotechnologie (Forschungszentrum Karlsruhe) is gratefully acknowledged.
MD and PN appreciate the support by the SFB~513 and granting of computer
time from the NIC, the SSC and the HLRS. 

\appendix

\section{Bulk force\label{sec:Bulk-force} }

We give here a short derivation of an estimate for the force needed
to break a bulk-like bond in a fcc lattice (cf.~Eq.~(\ref{eq:bulk_force_final})).
The reasoning follows Ref.~\onlinecite{RubioB2001} (see remark {[}25{]}
of that Ref.), where however no derivation is given.

The total energy of a crystal can approximately be written as $E_{N}(r)=NE(r)$,
where $N$ is the number of atoms in the volume $V$ of the considered
crystal, $E(r)=E_{\textrm{coh}}E^{*}(r^{*})$ is the energy of a single
atom as a function of the Wigner-Seitz radius, $E_{\textrm{coh}}$
is the equilibrium cohesive energy (or enthalpy of formation) and
$E^{*}(r^{*})$ is the {}``universal'' energy function $E^{*}(r^{*})=-\left(1+r^{*}\right)\exp \left(-r^{*}\right)$.\cite{Rose1998}
The Wigner-Seitz radius $r$ is defined as $r=\left(3/4\pi n_{A}\right)^{1/3}$
with the atom density $n_{A}$. Because $n_{A}=N/V=4/a^{3}$ in a
fcc crystal, $r$ is connected with the fcc-lattice constant $a$
via $r=\left(3/16\pi \right)^{1/3}a$. Additionally, $r^{*}=\left(r-r_{0}\right)/\ell $
is the scaled Wigner-Seitz radius and $r_{0}$ the equilibrium value
of $r$. The length scale $\ell $ is related to the bulk modulus
$B$, and it can be shown\cite{Rose1998} that \begin{equation}
\ell =\sqrt{\frac{E_{\textrm{coh}}}{12\pi Br_{0}}}=\left(\frac{16\pi }{3}\right)^{\frac{1}{6}}\sqrt{\frac{E_{\textrm{coh}}}{12\pi Ba_{0}}}\label{eq:bulk_screeninglength_deriv}\end{equation}
 with the equilibrium fcc-lattice constant $a_{0}$.

An estimate for the maximal force $F$ needed to break a bulk-like
bond may be obtained at the the inflection point of $E_{N}(r)$ at
a Wigner-Seitz radius $r_{\textrm{ip}}=r_{0}+\ell $ (where ip stands
for inflection point). If we use the relation $r=\left(3/16\pi \right)^{1/3}\sqrt{2}x$
between the Wigner-Seitz radius and the fcc-nearest-neighbor distance
(or interatomic bond length) $x$, an approximate bond length at rupture
of $x_{\textrm{ip}}=(a_{0}+\left(16\pi /3\right)^{1/3}\ell )/\sqrt{2}$
is obtained. The absolute value of the maximal force $F$ per bond
(where there are $6N$ bonds in a fcc lattice) is then given as\begin{equation}
F=\frac{1}{6N}\left.\frac{d\tilde{E}_{N}(x)}{dx}\right|_{x=x_{\textrm{ip}}}=\left(\frac{3}{16\pi }\right)^{\frac{1}{3}}\frac{\sqrt{2}E_{\textrm{coh}}}{6\exp \left(1\right)\ell }.\label{eq:bulk_force_deriv}\end{equation}
 (where $\tilde{E}_{N}(x)=E_{N}(r)$). This finally leads to the following
maximal force per bond in a fcc lattice:\begin{equation}
F=\sqrt{\frac{E_{\textrm{coh}}Ba_{0}}{8\exp \left(2\right)}}.\label{eq:bulk_force_final}\end{equation}
 In order to obtain numerical values from Eq.~(\ref{eq:bulk_force_final}),
we employ the data listed in Ref.~\onlinecite{Kittel1999} for the
three constants $E_{\textrm{coh}}$, $B$ and $a_{0}$.


\begin{thebibliography}{10}
\bibitem{remarkchargeCC}The circumstance that for simplicity matrices only extend over the
central part of our system and charges in the electrodes are neglected
has a slight influence on the conductance for thick contacts due to
interface effects. But during the stretching process, the narrowest
part of the junction plays an ever more important role, so that finally
this approximation becomes unimportant.
\bibitem{remarkMulliken}The Mulliken population analysis $N_{i}=\sum _{\alpha }(PS)_{i\alpha ,i\alpha }$
is chosen over the L\"{o}wdin analysis $N_{i}=\sum _{\alpha }(S^{1/2}PS^{1/2})_{i\alpha ,i\alpha }$,
which would be more consistent with the LDOS calculation (cf.~Eq.~(\ref{eq:LDOS_i_alpha_sigma}))
due to efficiency reasons. First the square root of the overlap does
not need to be computed in the Mulliken population analysis. But the
main advantage is the saving of an additional matrix multiplication
required in every step of the iterative solution of the charge-neutrality
condition, which requires that $|N_{i}-N_{\textrm{atom}}|<0.02$ for
every atom $i$ of the central part of the nanowire.
\bibitem{remarkTaverage}The averaged transmission $\left\langle T\right\rangle $ (cf.~Eq.~(\ref{eq:conductanceGav}))
is determined as a sum over $T(E)$ on $11$ equally spaced points
in the energy interval $[E_{F}-\Delta ,E_{F}+\Delta ]$ around the
Fermi energy ($\Delta =50$ meV).
\bibitem{remarketa}For plots of the LDOS of an atom a broadening of $\eta =10^{-3}\, \textrm{Ry}=0.0136\, \textrm{eV}$
is used in $\hat{G}_{CC,\sigma }$ (cf.~Eq.~(\ref{eq:DefGCCsigma})),
while for transport the broadening is chosen to be $\eta =10^{-8}\, \textrm{Ry}=1.36\cdot 10^{-7}\, \textrm{eV}$.
\bibitem{remarkchannelregions}For the division of a conductance trace into regions with different
numbers of open conductance channels we use a criterion of $T_{n}^{\sigma }<0.01$
to consider the $n$th channel to be closed. This division is only
approximate: Due to fluctuations in the geometry, a channel transmission
may fall temporarily below $0.01$ but reenter later on. In other
cases (especially Pt, Al and Ni) many channels are present and, in
order not to overload our pictures, we need to combine several channel-closing
events into one.
\bibitem{remarkPt-T1.5}Further investigations are needed to find out, why the transmission
of dimers and short chains is rather $1\, G_{0}$ than $1.5\, G_{0}$.
Possible reasons could lie in the structural properties, like for
example the high disorder of our thin contacts or linear instead of
zigzag chains. The approximation of surface Green's functions by bulk
Green's functions in the electrodes has been checked to have little
influence on the transmission when chains have formed. Potentially
the low transmission for dimers and short chains could also indicate
limitations for the use of the Pt TB parameters in conductance calculations.
\bibitem{remarkhalfquantization}More generally, the half-integer conductance quantization could also
arise from a perfectly polarized current, where the channel transmissions
of the transmitted spin-component add up to one.
\bibitem{remarkspinpolinset}Note that bins (where a of bin-width of $0.04\, G_{0}$ has been used)
containing only a single data point are not displayed in the inset
of Fig.~(\ref{cap:Ni-spinpolarization}). The reason is that the
calculation of the mean error of the arithmetic mean requires at least
two data points.\cite{remarkmeanchannel}
\bibitem{remarkbfhisto}The choice of the $20$ highest force values out of the last $30$
recorded geometries before the point of rupture for each of the $50$
simulated contacts can be justified as follows: To obtain enough statistics
on the breaking force, and because of the evident fluctuations in
the strain force (cf.~Fig (\ref{cap:Ag-contact2}), for example)
several geometries after breaking of a wire need to be considered
in the breaking force histogram. This is why we take into account
$30$ recorded geometries before the point of rupture. However, the
very tiny strain forces directly after and at the point of rupture
(which is signaled by a negligible MCS of less than $0.3\, \textrm{Å}$)
will produce artificial peaks at small force values. For this reason
we select the $20$ highest strain forces from the set of the $30$
strain force values. This is done for each of the $50$ simulated
contacts, and the data is complied in the histogram of the breaking
force.
\end{thebibliography}
\end{document}